\documentclass[%
reprint,
 amsmath,amssymb,amsfonts,
aps,
pra,
floatfix
]{revtex4-2}

\usepackage{graphicx, placeins}
\usepackage{dcolumn}
\usepackage{bm}

\begin{document}

\preprint{APS/123-QED}

\title{Weak-field expansion: A time-closed solution of quantum three-wave mixing}

\author{Hanzhong Zhang}
\author{Avi Pe'er}%
 \email{avi.peer@biu.ac.il}
\affiliation{%
Department of Physics and BINA Center for Nano-technology, Bar Ilan University, Ramat Gan 5290002, Israel
}%

\date{\today}

\begin{abstract}

We present a systematic derivation of the Heisenberg evolution of a trilinear bosonic Hamiltonian system in the presence of a strong drive beyond the standard approximation of a classical, undepleted driving field. We employ a perturbative expansion of the Hamiltonian propagator \textit{in orders of the input field amplitudes}, as opposed to the standard Baker-Campbell-Hausdorff (BCH) expansion of the propagator in orders of time. Our method automatically provides time-closed expressions and converges considerably faster than BCH expansion, especially in the regime of high parametric gain because the small parameter it uses is natural to the problem. We obtain the well-known quantum solution for optical parametric amplification of down-conversion simply as the first order of the expansion and present the rigorous procedure to derive higher-order corrections one by one. To demonstrate the utility of higher corrections, we discuss the second-order correction to the pump field as an \textit{ideal} detector of time-energy entanglement in parametric down-conversion. We also use the third-order correction to calculate the limits on the achievable squeezing with parametric amplification and on the fidelity of quantum-state transfer from one optical mode to another with sum (difference) frequency generation, due to the quantum properties of the strong driving field. 

\end{abstract}

\maketitle


\section{Introduction}
Three-wave mixing (TWM) is the lowest-order nonlinear couplings between modes that stand as a cornerstone of quantum science and technology.  This trilinear bosonic interaction structure appears whenever a strong coherent tone mediates exchange or pair creation between two weaker quantum modes. Examples include Josephson three-wave mixers in superconducting circuits, cavity optomechanical frequency conversion under strong pumping, electro-optic microwave–optical transduction, etc. \cite{Josephson, RevModPhys.86.1391, mw_transduction}. 

In quantum optics, TWM serves the generation, manipulation, and measurement of entangled or squeezed states of light \cite{SPDC1970, Polarization1995, EPR1988, Shaked2014} for applications of quantum communication, quantum-enhanced interferometric sensing, and photonic quantum computing \cite{Raymer_2020, Alon2026, 10.1063/5.0023103, Quantum_dots}. For example, both sum-frequency generation (SFG) and difference-frequency generation (DFG) can be used to transfer an arbitrary quantum state of light from one frequency to another \cite{QFC1990, Nature2005, QFC2012}. Other major examples of quantum wave-mixing are parametric down-conversion (PDC) and spontaneous PDC (SPDC), the leading sources of Einstein-Podolsky-Rosen entanglement and squeezed light \cite{epr, bell1964einstein, EPR1988, ekert1991quantum}, where the nonclassical correlation between the modes is driven by the production of photon pairs (namely, signal and idler) from a beam of strong coherent pump such that the frequency sum of each photon pair equals exactly the pump frequency (energy conservation).

The leading theoretical treatment of TWM in all the platforms above assumes the linearized or undepleted approximation, which provides a simple and powerful framework for quantum analysis of many practical scenarios. However, this treatment is \textit{inherently incomplete}, since it systematically discards the backaction of the weak modes onto the strong drive and omits corrections induced by the drive’s quantum fluctuations. Specifically, when analyzing TWM, one usually approximates the driving field as an infinitely strong coherent classical source that is unaffected by the nonlinear conversion, thereby simplifying the solution for the quantum evolution of the other fields. For example, in the quantum-state transfer based on SFG or DFG, we assume that the idler is an undepleted classical source, which leads to a periodic exchange of the quantum states between the signal and the pump (similar to Rabi flops in a two-level system) \cite{QFC1990, QFC2012, Nature2005, QObook}, according to
\begin{subequations}\label{previous_transfer}\begin{align}
\begin{split}
    \hat{a}_s(t)=\hat{a}_s(0)\cos{g}_i-\mathrm{i}\mathrm{e}^{-\mathrm{i}\phi_i}\hat{a}_p(0)\sin{g_i},
\end{split}\\
\begin{split}
    \hat{a}_p(t)=\hat{a}_p(0)\cos{g_i}-\mathrm{i}\mathrm{e}^{\mathrm{i}\phi_i}\hat{a}_s(0)\sin{g_i},
\end{split}
\end{align}\end{subequations}
where $t$ is the time and $g_i=\left|\alpha_i\right|\chi t$ is the imaginary gain of the oscillation, with $\alpha_i$ the classical complex amplitude of the strong idler and $\chi$ the reduced nonlinear coefficient. When $g_i=\pi/2$, the signal and the pump will exactly exchange their quantum states with $100\%$ fidelity. However, when one takes into account also the quantum fluctuations in the idler field, decoherence will kick in and the transfer fidelity will no longer be $100\%$. Such quantum imperfections are qualitatively expected, but a quantitative analysis of this effect is still lacking.

Similarly, when we take the assumption that the pump of a PDC process is undepleted, the evolution of the signal and idler can be solved analytically as \cite{QObook}
\begin{equation}\label{previous_PDC}
\hat{a}_{s,i}(t)=\hat{a}_{s,i}(0)\cosh{g_p}-\mathrm{i}\mathrm{e}^{\mathrm{i}\phi_p}\hat{a}_{i,s}^\dagger(0)\sinh{g_p},
\end{equation}
which is the standard description of an optical parametric amplifier (OPA), where the signal and idler fields are amplified via the parametric gain $g_p=|\alpha_p|\chi t$ in the nonlinear medium. This solution is usually sufficient to describe the quantum behavior of the down-converted fields, such as their nonclassical intensity and phase correlations, as well as the interference between the signal-idler pairs and the pump in multiple OPAs [known as SU(1,1) interference] \cite{PhysRevA.44.4614, PhysRevA.56.3214, Shaked2014}. Yet, although the assumption of undepleted drive provides a simple and intuitive approximate description of the dynamics, it also prevents deeper understanding of the quantum nature of the process, as the significant evolution of the pump is totally ignored. 

For example, in the case of broadband multimode PDC, the generated signal-idler photon pair is energy-time entangled. Standard methods to detect the entanglement provide only partial information (only time or only energy correlation, but not both), since any direct (local) measurement of the signal or idler photons in one dimension (e.g., time) collapses the entanglement and prevents observation of the correlation in the conjugate dimension (energy). However, the pair production or annihilation clearly modifies the pump field, alluding to the possibility to obtain information on the \textit{global} state of the photon pairs by measuring the pump, without local measurements of the individual signal or idler photons. Indeed, we demonstrated recently that coherent detection of the pump depletion or SFG can form a \textit{global} sensor of time-energy entanglement with superior signal-to-noise ratio compared to any technique based on local measurements \cite{us}.

Thus, analytic solutions beyond the undepleted field approximation are of great significance. The standard approach to consider the depletion is to apply the Baker-Campbell-Hausdorff (BCH) formula \cite{Hausdorff1906, DiracQM,Magnus1954, QObook}
\begin{equation}\begin{split}\label{bch}
    &\hat{a}(t)=\mathrm{e}^{{\mathrm{i}\hat{H}t}/{\hbar}}\hat{a}(0)\mathrm{e}^{-{\mathrm{i}\hat{H}t}/{\hbar}}\\
    &=\hat{a}(0)+\frac{\mathrm{i}t}{\hbar}[\hat{H},\hat{a}(0)]-\frac{t^2}{2\hbar^2}\bigl[\hat{H},[\hat{H},\hat{a}(0)]\bigr]+\cdots,
\end{split}\end{equation}
which expands the output operators in powers of the evolution time. Although the BCH expansion can produce the complete solution in principle, it is useful only for short interaction time, because for longer times the series converges very slowly, requiring an expansion to very high orders. One way to avoid long inconvenient expressions is to collect the leading terms according to the patterns of the commutators, as performed by Xing and Ralph \cite{PRA_pump_depletion}; yet determining the form of the solutions is not always easy, especially in the multimode case, where multiple signal-idler pairs of modes interact with the pump field (either a single mode or multiple modes).

Beyond the BCH-style expansions in time, other quantum treatments of parametric amplification with a quantized pump have addressed various corrections to the classical-pump approximation and established important limits on the achievable squeezing of a parametric amplifier imposed by the quantum fluctuations of the pump. Hillery and Zubairy used a coherent-state path-integral expansion of the propagator for the degenerate parametric amplifier \cite{HilleryZubairy1984}, while Crouch and Braunstein \cite{CrouchBraunstein1988}, Kinsler \textit{et al.} \cite{KinslerFerneeDrummond1993}, and Cohen and Braunstein \cite{CohenBraunstein1995} analyzed pump-noise and depletion effects mainly through semiclassical large-pump methods and positive-$P$ benchmarks. Similar to our approach, these prior works derived expansions of the Heisenberg evolution in powers of $1/N_p$ (the photon number of the strong pump field), but their expansion was drastically different. First, these works relied strongly on a semiclassical approximation of the quantum noise, where all operatorial expressions were replaced by c-number random functions. We instead utilize the algebraic properties of canonical operators to expand the Heisenberg evolution in a pure-quantum manner, retaining the parametric gain as a quantum operator function of the strong field. This allows us to obtain closed-form expressions in every order for all the output field operators in terms of the inputs. In addition, the applications in those works focus mainly on corrections to the quadrature variations (the squeezing), whereas we directly expand the quantum fields themselves, allowing us to calculate later any desired quantity in a general, fully quantum-mechanical manner. With our general quantum approach, we calculate the limits on squeezing in Sec.~\ref{squeezing_limit} and compare to the semiclassical results of \cite{CrouchBraunstein1988}. Our results highlight the implications of a quantum treatment of the pump phase and of the experimental choice of phase reference for evaluating the squeezing. Specifically, deriving the reference from the evolving pump field leads to a qualitatively different correction compared to a classical nonevolving reference, primarily due to the inclusion of commutation relations in the derivation of the field evolution, as we explain in Sec.~\ref{squeezing_limit}.

In addition, quantum treatments of three-wave mixing were pursued extensively within the setting of an optical parametric oscillator (OPO) cavity, where the trilinear interaction is embedded in a driven-dissipative intracavity master equation with Markovian coupling to external baths and is commonly analyzed using phase-space methods such as the positive-$P$ representation \cite{DrummondGardiner1980,PhysRevLett.60.2731,PhysRevA.41.3930}. Experimentally, OPOs above threshold have revealed strong continuous-variable correlations and bipartite or multipartite entanglement structures involving pump, signal, and idler fields \cite{PhysRevLett.95.243603,PhysRevLett.97.140504,Coelho2009ThreeColorEntanglement}. Yet our work here treats directly the unitary dynamics of the bare three-wave-mixing Hamiltonian in the nonlinear medium, without assuming any cavity or Markovian system-bath coupling; to analyze these scenarios, one may combine the medium propagator given by our solution with standard input-output modeling of the cavity.

The rest of this paper is organized as follows. We first describe the general concept of the weak-field expansion and how the dynamically evolving field operators can be analytically calculated order by order (Sec.~\ref{general_concept}). This general and complete expansion is not tied to a particular input state or density matrix, such as vacuum input in the weak modes, or a particular choice of the strong mode. It therefore automatically applies to any form of parametric amplification, i.e., seeded, cascaded with multiple OPAs in series, multimode, etc., as well as to quantum-state transfer via frequency conversion, as we present herein. We then show how the standard equations of an OPA with undepleted pump are derived from the first-order expansion (Sec.~\ref{First_order}) and then calculate the leading corrections in two major scenarios: parametric amplification and quantum-state transfer. We discuss the correction to the strong-field operator in Sec.~\ref{pump_change}, where we derive the second-order correction to the pump field of an OPA and point to its utility as a global detector of the signal-idler entanglement in Sec.~\ref{pump_change}. Section~\ref{third_order} derives the correction to the weak-field operators (third order), where Sec.~\ref{squeezing_limit} calculates the limit on the obtainable squeezing in comparison with the semiclassical calculation of \cite{CrouchBraunstein1988}, and Sec.~\ref{state_transfer} calculates the fundamental limit to the fidelity of quantum-state transfer. The Appendix dives into the details on the expansion procedure and reviews the mathematical techniques of the derivation. 

\section{Weak-Field Expansion: General Concept}\label{general_concept}
We introduce a complementary perturbative expansion that reorganizes the Heisenberg evolution of the trilinear Hamiltonian $\hat{H}\!=\!\hbar\chi\left(\hat{a}_p\hat{a}_s^\dagger\hat{a}_i^\dagger+\hat{a}_p^\dagger\hat{a}_s\hat{a}_i\right)$ by \textbf{orders of the weak-field operators} rather than by powers of time. We directly solve the time-closed evolution of each order analytically, yielding a controlled hierarchy whose leading order reproduces the standard linearized or undepleted-drive physics, while higher orders provide corrections that capture depletion and drive-induced quantum noise. In this sense, the method is analogous to the BCH expansion\textemdash yet with a perturbative parameter that is directly tied to the \textbf{excitation hierarchy} (strong vs weak fields)\textemdash the natural small parameter in a wide class of driven quantum systems. Such an expansion converges fast as long as the change in the population of the strong driving field (depletion or gain) is small compared to its original population along the whole process $\Delta N_{\mathrm{drive}}(t)\ll N_{\mathrm{drive}}$, indicating that the approximation is valid under arbitrarily high gain (and its validity improves with the inclusion of higher-order corrections).

The derivation is guided by the ansatz that the time evolution of the operator $\hat{a}(t)$ can be written as
\begin{equation}\label{Ansatz}
    \hat{a}(t)=\hat{a}^{(0)}(t)+\hat{a}^{(1)}(t)+\hat{a}^{(2)}(t)+\cdots,
\end{equation}
where each term $\hat{a}^{(k)}(t)$ in the series is a time-closed expression that reflects the sum of all the BCH terms with weak-field operators $a_w$ and $a_w^{\dagger}$ of power $k$. To calculate $\hat{a}^{(k)}(t)$, we plug the formal ansatz of Eq.~\ref{Ansatz} into the Heisenberg differential equations of the dynamical evolution for the signal, idler, and pump:
\begin{subequations}\label{general evolution}
\begin{align}
    &\frac{\mathrm{d}}{\mathrm{d}t}\hat{a}_p(t)=\frac{\mathrm{i}}{\hbar}\bigl[\hat{H},\hat{a}_p(t)\bigr]=-\mathrm{i}\chi \hat{a}_s(t)\hat{a}_i(t),\\
    &\frac{\mathrm{d}}{\mathrm{d}t}\hat{a}_{s,i}(t)=\frac{\mathrm{i}}{\hbar}\bigl[\hat{H},\hat{a}_{s,i}(t)\bigr]=-\mathrm{i}\chi\hat{a}_{i,s}^\dagger(t)\hat{a}_p(t).
\end{align}
\end{subequations}

We start the derivation by considering the zeroth-order solution, which keeps terms with only $\hat{a}_{s,i,p}^{(0)}(t)$ in Eq.~(\ref{general evolution}); this is equivalent to the trivial classical solution, which completely omits the contributions of any weak field or quantum effect. The first-order correction is derived by a perturbative solution of Eq.~\ref{general evolution}, assuming formally a solution of the form $\hat{a}_x^{(0)}(t)+\hat{a}_x^{(1)}(t)$, keeping terms up to first order in the input fields. This calculation can then be repeated up to any order: If we obtain the solutions for all the leading $k$ orders, we can solve the $(k+1)$th order, thereby generating the complete solution of the Heisenberg dynamics order by order. This is the practical origin of the rapid convergence of the weak-field expansion in regimes where the fractional change of the strong-field population remains small, even when the parametric gain is large. As we will see from the derivation below, the expansion of orders provides an immediate bookkeeping interpretation: The even orders account for the dynamics of the strong field, whereas the odd orders affect the weak-field dynamics. Thus, strong-field corrections, such as pump depletion and pump-induced quantum noise, enter through the \emph{even} orders, while the weak fields acquire their leading quantum evolution through the \emph{odd} orders. A detailed description of the general procedure is provided in the Appendix.

To exemplify this procedure, we analyze in depth two canonical scenarios of quantum information with TWM and derive the leading quantum correction beyond the undepleted assumption. First, we solve parametric amplification or attenuation with a strong pump [Eq.~(\ref{previous_transfer})] and general inputs up to the third order, where we uncover the quantum dynamics of the pump in squeezing, SU(1,1) interference, etc., the utility of the pump as a global detector of entanglement with superior rejection of noise or decoherence, and derive the squeezing limit for an ideal OPA in comparison with the semiclassical result. Second, we examine the transfer of a quantum state from one wavelength to another using sum-difference frequency generation with a strong drive [Eq.~(\ref{previous_PDC}), e.g., to transfer quantum light from the IR range to the visible, where efficient detectors are available to process the quantum information]. We calculate the third-order correction for the transfer and derive the fidelity limit due to the quantum fluctuations of the strong drive.

\section{First-order Solution: standard undepleted pump}\label{First_order}

Let us first consider parametric amplification, i.e., TWM with a strong pump and weak signal and idler. The zeroth-order solutions are $\hat{a}_p^{(0)}(t)\!\equiv\! \hat{a}_p(0)$ and $\hat{a}_{s,i}^{(0)}(t)\!\equiv\!0$, indicating that classical PDC cannot occur spontaneously without seeding. Following the procedure described above, the first-order equations are
\begin{subequations}\label{1st order equations}\begin{align}
\begin{split}
    \frac{\mathrm{d}}{\mathrm{d}t}\hat{a}_p^{(1)}(t)=\mathrm{i}\chi\hat{a}_{s}^{(0)}(t)\hat{a}_i^{(0)}(t)=0,
\end{split}\\
\begin{split}
    \frac{\mathrm{d}}{\mathrm{d}t}\hat{a}_{s,i}^{(1)}(t)=-\mathrm{i}\chi\hat{a}_{i,s}^{(1)\dagger}(t)\hat{a}_p^{(0)}(t).
\end{split}
\end{align}\end{subequations}
Considering the initial conditions for the operators $\hat{a}_p^{(1)}(0)=0$ and $\hat{a}_{s,i}^{(1)}(0)=\hat{a}_{s,i}$,
the solutions to Eq.~\ref{1st order equations} are
\begin{subequations}\label{1st order solutions}
\begin{gather}
\begin{split}
    \hat{a}_p^{(1)}(t)\equiv 0,
\end{split}\\
\begin{split}
    \hat{a}_{s,i}^{(1)}(t)=\hat{a}_{s,i}\cosh{\hat{g}_p}-\frac{\mathrm{i}\hat{a}_{i,s}^\dagger\hat{a}_p}{\hat{N}_p^{1/2}}\sinh{\hat{g}_p},
\end{split}
\end{gather}
\end{subequations}
where $\hat{N}_p=\hat{a}_p^\dagger\hat{a}_p$ is the pump photon-number operator and $\hat{g}_p=\hat{N}_p^{1/2}\chi t$ is the gain operator. On the right-hand side of Eq.~\ref{1st order solutions}b, we follow the convention that field and number operators are taken as the input operators $\hat{a}=\hat{a}(0)$ and $\hat{N}=\hat{a}^\dagger(0)\hat{a}(0)$; this convention is adopted throughout our analysis (unless the operators are directly specified as time dependent). 

Equation~(\ref{1st order solutions}b) is almost the same as Eq.~(\ref{previous_PDC}), the well-known OPA equation with a classical undepleted pump, except for the operator representation of the pump field (number operator $\hat{N}_p$ in the gain $\hat{g}_p$). This is necessary for correctly calculating the higher-order solutions and allows calculations on the quantum fluctuations of the strong pump field beyond the classical assumption of $\hat{a}_p\approx\alpha_p$ and $\hat{N}_p^{1/2}\approx\left|\alpha_p\right|$. Additionally, note that the square-root representation $\hat{N}_p^{1/2}$ and the operator in the denominator are actually well defined, since Eq.~\ref{1st order solutions}b can be expanded into a power series of $\hat{N}_p$ as $\hat{N_p}^{-1/2}\sinh{\hat{g}_p}\!\approx\!\chi t\!+\!\hat{N}_p{(\chi t)}^3/3!\!+\!\hat{N}_p^2{(\chi t)}^5/5!\!+\!\cdots$, corresponding to the complete bookkeeping of the leading terms in the BCH formula.

Yet, under the new principles, the operators $\hat{a}_{s,i}^{(1)}$ and $\hat{a}_p^{(0)}(t)$ will no longer commute with each other (since $\hat{a}_{s,i}^{(1)}$ contains a component of $\hat{N}_p$), and Eq.~\ref{1st order solutions} will not be the exact solution to Eq.~\ref{1st order equations}. However, this error will affect only the third order or higher (as explained in Appendix~\ref{principle details}) and will be accounted for in the derivation of the next orders.

\section{Second-order Solution: Correction to the strong field}\label{pump_change}
We can now continue the derivation to the second order. The perturbative equations are
\begin{subequations}\label{2nd order equations}
\begin{align}
\begin{split}
    \frac{\mathrm{d}}{\mathrm{d}t}\hat{a}_p^{(2)}(t)=-\mathrm{i}\chi\hat{a}_s^{(1)}(t)\hat{a}_i^{(1)}(t),
\end{split}\\
\begin{split}
    \frac{\mathrm{d}}{\mathrm{d}t}\hat{a}_{s,i}^{(2)}(t)=-\mathrm{i}\chi\hat{a}_p^{(0)}(t)\hat{a}_{i,s}^{(2)\dagger}(t). \end{split}
\end{align}
\end{subequations}
For the weak signal-idler field, the second-order correction is nulled $\hat{a}_{s,i}^{(2)}(t)\!\equiv\! 0$ because Eq.~(\ref{2nd order equations}b) is homogeneous with a null initial condition $\hat{a}_{s,i}^{(2)}(0)\!=\!0$. The pump correction is
\begin{equation}\begin{aligned}\label{pump depletion}
    \hat{a}_p^{(2)}(t)\!=\!-\frac{\mathrm{i}\hat{a}_s\hat{a}_i}{2\hat{N}_p^{1/2}}&\sinh{2\hat{g}_p}\!-\!\hat{a}_p\frac{\hat{N}_s\!+\!\hat{N}_i+1}{2\hat{N}_p}\sinh^2{\hat{g}_p}\\
    +\frac{\mathrm{i}\hat{a}_p\hat{H}'}{4\hat{N}_p^{3/2}}&(\sinh{2\hat{g}_p}\!-\!2\hat{g}_p),
\end{aligned}\end{equation}
where $\hat{H}'\!=\!\hat{a}_p^\dagger\hat{a}_s\hat{a}_i\!+\!\hat{a}_p\hat{a}_s^\dagger\hat{a}_i^\dagger$ is the unitless Hamiltonian. 

Let us now consider the physical consequences of the second-order correction. Equation~(\ref{pump depletion}) above decomposes the correction to the pump field into three terms, whose interference dictates its amplitude and phase. Let us elaborate on the physical meaning of the three terms for some intuition. (i) The first term presents the possibility to recombine the signal-idler pairs into the original pump via SFG, which reflects time reversal of the PDC process. This is the only term that relies on the coincidence of the signal and the idler photons, depending on their phase. Accordingly, the pump response to the signal-idler input (depletion or SFG) can act as an ideal detector of photon pairs, especially in the case where the monochromatic pump generates a multimode spectrum of down-conversion, as was recently demonstrated \cite{us} [we will elaborate on this further in the discussion of Eq.~(\ref{2nd multimode})]. (ii) The second term always has an opposite phase to the original pump and is linearly dependent on the total intensity of the signal and idler input, indicating a direct depletion of the pump due to the generation of the signal-idler photons. (iii) The last term is always in quadrature to the input pump and therefore reflects a nonlinear phase shift (similar to a Kerr phase). Since this last term shows a third-order dependence on the gain $\hat{g}_p$, it is negligible at the low-gain regime, whereas for sufficiently large gain values, it may become comparable to the first two (or even dominate them) and should also be taken into account.

While this breakdown of terms is only an intuitive interpretation of the mathematical solution and each term cannot be detected independently, the signal-idler coincidence in the first term indicates the possibility to utilize the pump generation or depletion as a global sensor of entanglement.
\begin{figure}
    \centering
    \includegraphics[width=1 \linewidth]{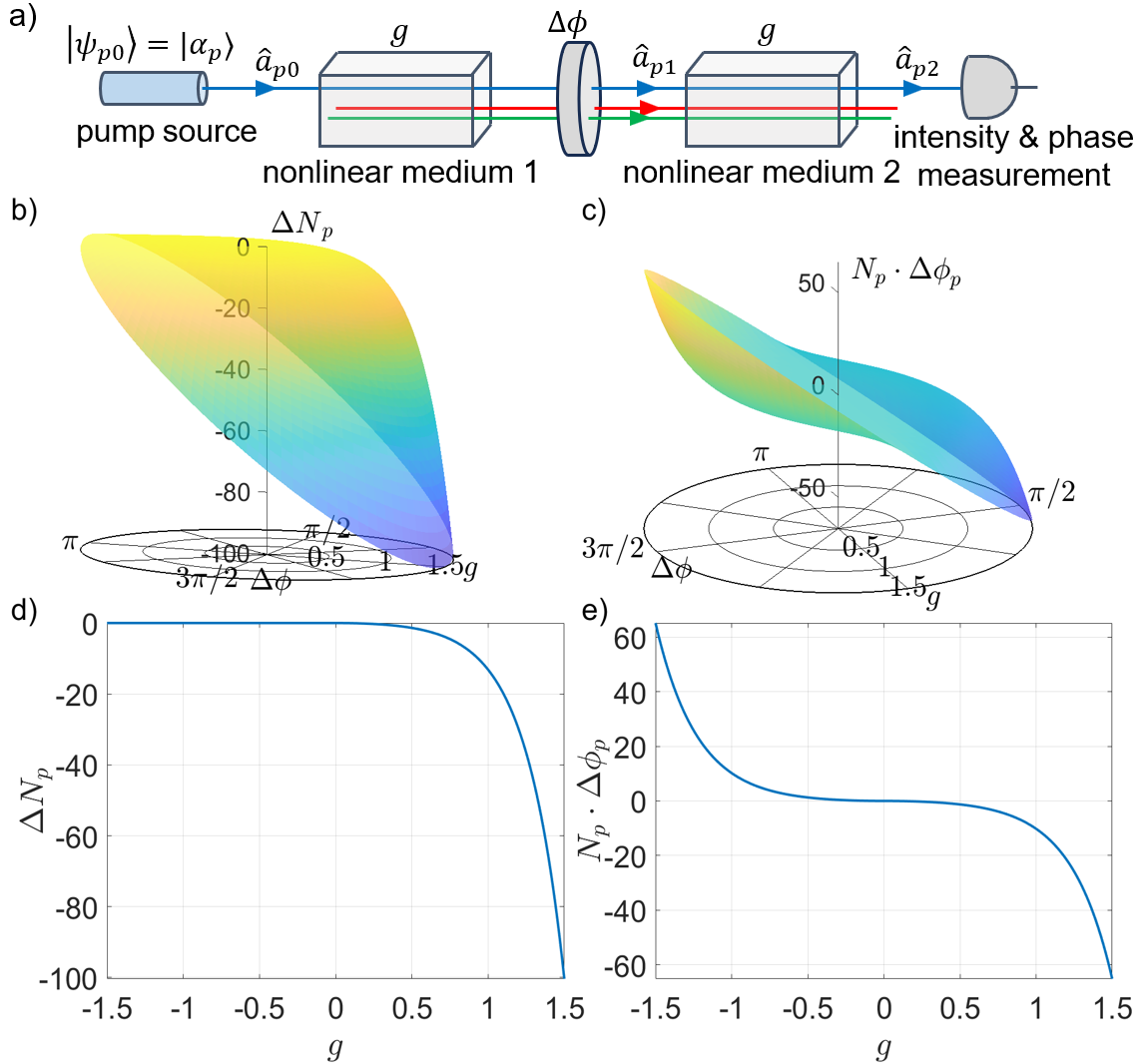}
    \caption{Second-order correction to the pump field at the output of an SU(1,1) interferometer. (a) SU(1,1) concept with marking of the pump field at the input and output. (b) Number of photons depleted or added to the pump $\Delta N_p\!=\!\bigl\langle\hat{a}_{p2}^{\dagger}\hat{a}_{p2}\bigr\rangle\!-\!|\alpha_{p}|^2$ due to PDC or SFG. (c) Phase of the output pump field $\Delta\phi_p$, as a function of the parametric gain $g$ (radius) and the total phase of the input fields $\Delta\phi$ (angle), plotted in polar coordinates. Also shown are section plots (d) along the $\pi$-$0$ phase line (real axis) of (a) and (e) along the $3\pi/2$-$\pi/2$ phase line (imaginary axis) of (b) (negative $g$ on the $x$ axis indicates a $\pi$ phase). We assume an ideal SU(1,1) interferometer (no internal loss) with equal parametric gain $g$ in the two nonlinear media.}
    \label{Fig: Pump change}
\end{figure}

\noindent\textbf{Global sensing of entanglement through pump measurement}

Equation~(\ref{pump depletion}) allows us to calculate the pump dynamics through a complete SU(1,1) interferometer [see Fig.~\ref{Fig: Pump change}(a)], where a single coherent pump laser drives two consecutive ideal OPAs, with a phase shift $\Delta\phi$ in between. If we measure the change of the pump field after the second OPA, in either intensity or phase, we will be able to detect the annihilation or generation of entangled signal-idler pairs in that second OPA without measuring any of the individual signal or idler photons. This forms a new type of global sensor of time-energy entanglement with superior noise rejection compared to local measurements, as demonstrated in \cite{us}. The remainder of this section provides the theory for using the pump as a quantum sensor. 

Equation~(\ref{pump depletion}) predicts the Heisenberg evolution of the pump field as it propagates through each of the nonlinear media and interacts with the signal-idler field in that medium ($\hat{a}_{p0}\rightarrow\hat{a}_{p1}$ in the first medium and $\hat{a}_{p1}\rightarrow\hat{a}_{p2}$ in the second). Figures~\ref{Fig: Pump change}(b)--\ref{Fig: Pump change}(e) depict the photon depletion $\Delta N_p\!=\!\langle N_{p2}\!-\!N_{p0}\rangle$ and phase shift $\Delta\phi_p=\!\mathrm{i}(\langle\hat{a}_{p2}\rangle/\alpha_p\!-\!\langle\hat{a}_{p2}^\dagger\rangle/\alpha_p^*)/2$ of the pump at the output of the SU(1,1) interferometer relative to the input as a function of the parametric gain $g\!=\!|\alpha_p|\chi t\!=\!{\bigl\langle\hat{N}_p\bigr\rangle}^{1/2}\chi t$ and the total phase $\Delta\phi\!=\!\phi_s\!+\!\phi_i\!-\!\phi_p$. When $\Delta\phi=\pi$ [destructive SU(1,1) interference], the standard undepleted pump analysis predicts for the second medium complete annihilation of all the biphoton pairs that were generated in the first medium. In accordance, our calculated pump correction preserves exactly the pump intensity at the SU(1,1) output ($\Delta N_p=0$ at $\Delta\phi=\pi$) for any value of $g$, indicating that the SFG in the second nonlinear medium exactly compensates for the pump depletion in the first medium, consistent with the Manley-Rowe conservation of photons in TWM. Thus, detection of the change in the photon number of the pump directly indicates the number of entangled signal-idler pairs that were annihilated or generated, providing a global coincidence detector that detects the pair as a whole, without local detection of each of the constituent photons. Furthermore, detection of the phase of the output pump allows us to determine the phase sum of the incoming entangled biphotons without knowing their individual input phases, which can be achieved by a simple reverse mapping. 
\begin{figure}
    \centering
    \includegraphics[width=1\linewidth]{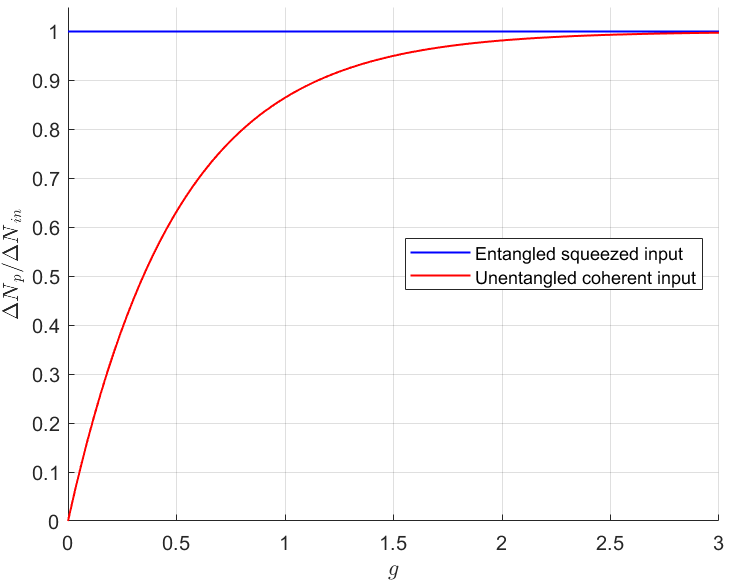}
    \caption{Nonclassical nature of SFG with squeezed input. We compare the efficiency of stimulated SFG $\Delta N_p/N_{in}$ (the number of SFG photons that are added to the pump relative to the number of input signal and idler photons) as a function of parametric gain $g$, under two input conditions: (i) ideally squeezed input (blue), as generated by SPDC with the same gain, i.e., SU(1,1) interference at the $\pi$ phase, and (ii) equal unentangled signal and idler (red), an upper limit for the conversion efficiency with coherent quantum input, or the classical prediction of stimulated SFG. Evidently, the SFG with an entangled input remains ideal for any gain $g$, while for a coherent unentangled input the SFG efficiency vanishes for low $g$, as classically expected.}
    \label{Fig: comparison}
\end{figure}

The signal-idler entanglement within the SPDC field plays a critical role for the SFG process. Figure~\ref{Fig: comparison} shows the SFG efficiency $\eta_{SFG}=\Delta N_p/N_{s,i}$ as a function of the parametric gain $g$ for an entangled input vs the efficiency for coherent unentangled signal and idler for the same parametric gain (classical efficiency). The figure highlights the quantum advantage of entanglement for SFG in the low-gain regime (single photon pairs), where the quantum SFG efficiency remains ideal, whereas the coherent SFG efficiency drops to zero. This advantage is the drive for applications of quantum sensing with entangled photons, such as a quantum radar or quantum illumination \cite{us, liu2020joint, PhysRevLett.118.040801}, where the entanglement dramatically enhances the wanted signal-idler SFG compared to the unentangled background, thereby forming a quantum matched filter with superior noise rejection than the best classical matched filter \cite{us}.

Let us now generalize our method to the case of a multimode spectrum of down-converted signal-idler pairs that are generated by a monochromatic single-mode pump. The solutions again can be derived from the same analysis. We can solve the first-order equations of each mode separately, just like the single-mode case, so Eq.~(\ref{1st order solutions}b) holds valid, and the second-order correction to the pump simply becomes a sum over all the individual contributions of the interacting modes:
\begin{equation}\label{2nd multimode}\begin{split}
\hat{a}_p^{(2)}(t)\!=\!-\frac{\mathrm{i}\hat{A}}{2\hat{N}_p^{1/2}}\sinh{2\hat{g}_p}-\hat{a}_p\frac{\hat{N}_\mathrm{dc}}{2\hat{N}_p}\sinh^2{\hat{g}_p}\\
+\frac{\mathrm{i}\left(\hat{A}\hat{N}_p+\hat{A}^\dagger\hat{a}_p^2\right)}{4\hat{N}_p^{3/2}}(\sinh{2\hat{g}_p}-2\hat{g}_p),
\end{split}\end{equation}
where
\begin{align}\label{A & N_dc}
\begin{split}
    &\hat{A}=\sum_{\omega_s+\omega_i=\omega_p}{\hat{a}_s\hat{a}_i},\\
    &\hat{N}_\mathrm{dc}=\sum_{\omega_s+\omega_i=\omega_p}{\left(\hat{N}_s+\hat{N}_i+1\right)},
\end{split}
\end{align}
and we assume the same parametric gain $\hat{g}_p$ for all modes. 

This multimode result is very important to the analysis of time-energy entanglement. It illuminates the relative dynamics between the single-mode pump output and the broadband multimode input of the photon pairs. Coherent detection of the pump field (amplitude and phase) in this single mode allows us to measure both the time difference and energy sum of the entangled photon pair simultaneously without measuring either the time or the energy of the individual photons. In contrast, a direct local measurement of each single photon in the pair will force either the time or energy information to collapse \cite{us}. Additionally, the shot noise of the pump does not limit the observation of the time-energy entanglement, even when each signal-idler mode is occupied by considerably less than one photon, as the total number of SFG photons results from the coherent summation over all the signal-idler frequency pairs.

{\section{Third-order solution: Correction to the weak fields}}\label{third_order}

We can now calculate the first meaningful (third-order) corrections for the weak down-converted fields beyond the standard OPA solution of undepleted pump [Eq.~(\ref{1st order solutions}b)]
 \begin{equation}\label{sol_3rd}\begin{split}
  \hat{a}_{s,i}^{(3)}(t)=&-\frac{1}{4}\hat{a}_{i,s}^\dagger\hat{A}\frac{C_+(\hat{g}_p)}{\hat{N}_p}-\frac{\mathrm{i}}{4}\hat{a}_{s,i}\hat{a}_p^\dagger\hat{A}\frac{S_+(\hat{g}_p)}{\hat{N}_p^{3/2}}\\
  &+\frac{1}{4}\left(\hat{a}_{i,s}^\dagger\hat{a}_p^2\hat{A}^\dagger-\hat{a}_{s,i}\hat{N}_\mathrm{dc}\hat{N}_p\right)\frac{C_-(\hat{g}_p)}{\hat{N}_p^2}\\
   &+\frac{\mathrm{i}}{4}\left(\hat{a}_{i,s}^\dagger\hat{a}_p\hat{N}_\mathrm{dc}+\hat{a}_{s,i}\hat{a}_p\hat{A}^\dagger\right)\frac{S_-(\hat{g}_p)}{\hat{N}_p^{3/2}}
 \end{split}\end{equation}
(see the calculation in Appendix~\ref{derivation_3rd}), where $\hat{A}$ follows the definition of Eq.~(\ref{A & N_dc}) and
\begin{subequations}\begin{align}\label{CpmSpm}
\begin{split}
&C_\pm(\hat{g}_p)=\cosh^3{\hat{g}_p}-\cosh{\hat{g}_p}\pm \hat{g}_p\sinh{\hat{g}_p},\\    &S_\pm(\hat{g}_p)=\sinh^3{\hat{g}_p}\pm\left(\hat{g}_p\cosh{\hat{g}_p-\sinh{\hat{g}_p}}\right).
\end{split}
\end{align}\end{subequations}
We remind the reader that the derivation of the first-order solution neglected commutation relations like $\Bigl[\hat{a}_s^{(1)},\hat{a}_p^{(0)}\Bigr]$, since they contribute only to the third order or higher. Now that we are considering the third-order contribution, we should be aware of the sequential order of these operators in the first-order expressions. For convenience, we purposely order the operators in Eq.~(\ref{1st order solutions}b) as $\hat{a}_p\sinh{\hat{g}_p}$ instead of $\sinh{\hat{g}_p}\;\hat{a}_p$, which yields the simplest expressions for the third order. We can still neglect the commutation relation of the pump field in higher orders, like in $\Bigl[\hat{a}_s^{(3)},\hat{a}_p^{(0)}\Bigr]$, since it contributes only to the fifth order or higher. If one wishes to further calculate the fifth order or more, they can again choose the ordering of the operators within the third order [Eq.~(\ref{sol_3rd})] to simplify the expressions of the fifth order, and so on.

\subsection{Limits on squeezing}\label{squeezing_limit}
Equation~(\ref{sol_3rd}) can now be used to calculate the correction of order $1/\hat{N}_p$ to desired measurable quantities of a parametric amplifier, like the squeezing level. Specifically, we can calculate the variance of the squeezed quadrature (starting from vacuum inputs for the signal and idler) of an ideal parametric amplifier with a very high gain, which Eqs.~(\ref{previous_PDC}) and (\ref{bch}) cannot predict. This limit was previously calculated using semiclassical methods \cite{CrouchBraunstein1988} and the result was
\begin{align}\label{quad_old}
    \begin{split}
    &\bigl\langle|\Delta\hat{X}_-^{(\mathrm{sc})}(t)|^2\bigr\rangle=\frac{1}{2}\bigl\langle\hat{X}_-^{(\mathrm{sc})\dagger}(t)\hat{X}_-^{(\mathrm{sc})}(t)+c.c.\bigr\rangle\\
    &\quad=\frac{1}{4}\mathrm{e}^{-2g_p}+\frac{1}{{|\alpha_p|}^2}\biggl[\frac{1}{64}\left(\mathrm{e}^{2g_p}-\mathrm{e}^{-2g_p}\right)\\
    &\qquad-\frac{g_p}{8}+\frac{2g_p^2-3g_p}{16}\mathrm{e}^{-2g_p}+\frac{1}{16}\left(1-\mathrm{e}^{-4g_p}\right)\biggr]
\end{split}
\end{align}
Here the squeezed quadrature was defined as $\hat{X}_-^{(\mathrm{sc})}=\mathrm{i}\left(\hat{a}_s-\hat{a}_i^\dagger\right)/2$, which fixes the phase reference classically at $\phi_{p}\!=\!\pi/2$. The input pump was assumed to be in a strong coherent state $|\alpha_p\rangle$, with evolving (semiclassical) fluctuations, but the parametric gain was taken as a classical quantity of $g_p = \alpha_p\chi t$. In contrast, our quantum analysis allows the pump field to evolve quantum mechanically, including its phase, and the parametric gain remains a quantum operator. We can therefore examine the validity of this result under more general conditions than the semiclassical treatment allows.

Critically, we need to consider that defining a quadrature requires a reference phase from the pump or an external local oscillator, and the definition above assumes that reference to simply be $\pi/2$, or more rigorously, to choose the exponential phase factor $\mathrm{e}^{\mathrm{i}\phi}=\mathrm{i}$. However, the ``quantum phase'' of a field is not a well-defined concept in quantum mechanics. In a fully quantum analysis, the phase reference is taken from a specific optical field, and the quantum phase factor should represent the dynamical evolution of that field in time. To improve the definition, we can express the phase as $\mathrm{e}^{\mathrm{i}\phi_p}\rightarrow\hat{a}_p\hat{N}_p^{-1/2}$ or as $\mathrm{e}^{\mathrm{i}\phi_p}\rightarrow\hat{N}_p^{-1/2}\hat{a}_p$ (or a symmetrized combination of both). Quantum commutation causes slight differences between these definitions, highlighting the fundamental ambiguity of the phase operator in quantum mechanics. Although these differences do not disappear, even for large numbers of photons \cite{PhysicsPhysiqueFizika.1.49}, their implication on the calculated squeezing is small.

While a theoretical resolution of the quantum ambiguity is beyond the scope of our discussion here, it is evident that a classical c-number treatment of the phase factor ignores the time evolution of the reference field, as well as the commutation relations. We therefore consider another operational choice of the phase factor that is guided by the experimental realization of an SU(1,1) interferometer
\begin{equation}\label{td_phase_factor}
    \hat{X}_-^{(\mathrm{q})}(t)=\frac{\mathrm{i}}{2}\left[\hat{a}_s(t)+\mathrm{i}\hat{a}_i^\dagger(t)\hat{a}_p(t){\hat{N}_p^{-1/2}(t)}\right].
\end{equation}
Here the phase reference is the evolving pump itself (in a parametric homodyne scenario for the measurement of the squeezing). This dynamical definition of the quadrature retains both the quantum properties of the pump and the commutativity between the phase factor and the idler operator at any time $t$ during the evolution (in contrast, defining the phase factor with only the input operator of the pump will not commute with the evolved signal or idler). The definition of Eq.~(\ref{td_phase_factor}) yields our pure-quantum limit on the squeezed variations 
\begin{align}\begin{split}\label{fq quantum squeezing}
    &|\Delta\hat{X}_-^{(\mathrm{q})}(t)|^2=\hat{X}_-^{(\mathrm{q})\dagger}(t)\hat{X}_-^{(\mathrm{q})}(t)\\
    &=\frac{1}{4}\mathrm{e}^{-2\hat{g}_p}-\frac{1+n_m}{\hat{N}_p}\biggl[\frac{\hat{g}_p}{8}\mathrm{e}^{-2\hat{g}_p}+\frac{1}{32}(\mathrm{e}^{-4\hat{g}_p}-1)\biggr].
\end{split}\end{align}
where $n_m$ denotes total number of modes (assuming all modes are nondegenerate). Note that we keep the pump intensity and parametric gain as operators here; the expectation value of this quantity under coherent pump input can be calculated with the help of an identity 
\begin{align}\begin{split}\label{coherent average}
    \bigl\langle\alpha\big| f{(\hat{N})}\big|\alpha\bigr\rangle=f\bigl({|\alpha|}^2\bigr)&+\frac{1}{2}{|\alpha|}^2f''\bigl({|\alpha|}^2\bigr)\\
    &+O\Bigl[f\bigl({|\alpha|}^2\bigr)\cdot|\alpha|^{-4}\Bigr],
\end{split}\end{align}
where $f$ is an arbitrary smooth function of the number operator $\hat{N}$ (see the full derivation in Appendix~\ref{coherent_eval}). Note that the second term in the expression is actually of order $f\bigl(|\alpha|^2\bigr)\cdot|\alpha|^{-2}$ due to the second-order derivative; every derivative of $f$ with respect to ${|\alpha|^2}$ will increase the order (counted in $1/|\alpha|$) of the expression by $2$.

The correction to the squeezing level up to $1/|\alpha_p|^2$ yields
\begin{align}\label{var_quad}
\begin{split}
  \Bigl\langle{\big|\Delta\hat{X}_-^{(q)}(t)\big|}^2&\Bigr\rangle= \frac{\mathrm{e}^{-2g_p}}{4}+\frac{n_m\!+\!1}{32|\alpha_p|^2}\\+\frac{\mathrm{e}^{-2g_p}}{{|\alpha_p|}^2}&\biggl[ \frac{2g_p^2\!-\!(2n_m\!+\!1)g_p}{16}-\frac{n_m+1}{32}\mathrm{e}^{-2g_p}\biggr],
\end{split}\end{align}
where we applied Eq.~(\ref{coherent average}) to calculate the averages of $\hat{g}_p$ and $1/\hat{N}_p$ in Eq.~(\ref{fq quantum squeezing}). Taking $n_m=1$, the quantum result of Eq.~(\ref{var_quad}) reproduces part of Crouch and Braunstein's semiclassical result, but several terms disappear ($\mathrm{e}^{2g}/64$, $-\mathrm{e}^{-2g}/64$, and $-g_p/8$), which physically indicates that the limit to the squeezing at high gain is constant ($1/|\alpha_p|^2$) in our quantum analysis, and not exponentially growing ($\mathrm{e}^{2g}/64|\alpha_p|^2$), as the semiclassical result of \cite{CrouchBraunstein1988} predicts. 

To identify the physical source of this difference, we calculate the same correction with the semiclassical definition for the squeezed quadrature $\hat{X}_-^{(\mathrm{sc})}$ with a fixed classical phase reference:
\begin{align}\begin{split}\label{scq_difference}
    \bigl\langle|\Delta\hat{X}_-^{(\mathrm{sc})}(t)|^2\bigr\rangle=\bigl\langle|\Delta\hat{X}_-^{(\mathrm{q})}(t)|^2\bigr\rangle+\frac{\sinh{2g_p}-4g_p}{32{|\alpha_p|}^2},
\end{split}\end{align}
where the missing terms exactly reappear. Hence, when quantum corrections are targeted, the specific choice of the phase reference becomes critical for both the calculation and the experiment (a carefully designed experiment may be able to differentiate the evolving pump phase from that of a purely classical reference), and the c-number approximation in the semiclassical calculation may miss some of the quantum commutators, depending on the exact reference choice. 

\subsection{Fidelity limits of quantum-state transfer}\label{state_transfer}
Let us now examine the common application of transferring a quantum state from one optical frequency to another using SFG or DFG with a strong idler. Here the third order provides the leading correction that sets the fundamental limit on the fidelity of the transfer due to the quantum properties of the strong-driving field.

Assuming a strong idler with weak signal and pump at the input, the first- and second-order contributions are
\begin{subequations}\label{State_transfer_1_2}\begin{align}
    \hat{a}_s^{(1)}(t)=&\quad\hat{a}_s\cos{\hat{g}_i}-\frac{\mathrm{i}\hat{a}_p}{\hat{N}_i^{1/2}}\sin{\hat{g}_i}\;\hat{a}_i^\dagger,\\
    \hat{a}_p^{(1)}(t)=&\quad\hat{a}_p\cos{\hat{g}_i}-\frac{\mathrm{i}\hat{a}_s\hat{a}_i}{\hat{N}_i^{1/2}}\sin{\hat{g}_i},\\
\begin{split}
    \hat{a}_i^{(2)}(t)=&-\frac{\mathrm{i}\hat{a}_s^\dagger\hat{a}_p}{2\hat{N}_i^{1/2}}\sin{2\hat{g}_i}+\hat{a}_i\frac{\hat{N}_p-\hat{N}_s}{2\hat{N}_i}\sin^2{\hat{g}_i}\\
    &+\frac{\mathrm{i}\hat{a}_i\hat{H}'}{4\hat{N}_i^{3/2}}\bigl(\sin{2\hat{g}_i}-2\hat{g}_i\bigr),
\end{split}
\end{align}\end{subequations}
where the imaginary gain $\hat{g}_i$ due to the strong idler is again an operator. As expected, the first-order solution shows the standard Rabi-like equations of state transfer [Eq.~(\ref{previous_transfer})] that predict a complete exchange of the quantum state between the signal and the pump when the idler is classical with $g_i=\pi/2$. However, if we take the high-order corrections into account, the efficiency and fidelity of the transfer become limited, as we will show in the following.

Taking the multiplication in the first order as in Eqs.~(\ref{State_transfer_1_2}a) and (\ref{State_transfer_1_2}b) (the necessity of this claim is demonstrated in Sec.~\ref{pump_change}), the third-order corrections to the signal and pump are
\begin{subequations}\label{State_transfer_3}\begin{align}
\begin{split}
    \hat{a}_s^{(3)}(t)=-&\frac{1}{4}\hat{B}\hat{a}_p\frac{c_-(\hat{g}_i)}{\hat{N}_i}-\frac{\mathrm{i}}{4}\hat{a}_s\hat{a}_i\hat{B}\frac{s_-(\hat{g}_i)}{\hat{N}_i^{3/2}}\\
    +&\frac{1}{4}\left(\hat{a}_p\hat{a}_i^{\dagger2}\hat{B}^\dagger+\Delta{\hat{N}}\hat{a}_s\hat{N}_i\right)\frac{c_+(\hat{g}_i)}{\hat{N}_i^2}\\
    +&\frac{\mathrm{i}}{4}\left(\hat{a}_p\hat{a}_i^\dagger\Delta\hat{N}+\hat{a}_s\hat{a}_i^\dagger\hat{B^\dagger}\right)\frac{s_+(\hat{g}_i)}{\hat{N}_i^{3/2}},
\end{split}\\
\begin{split}
    \hat{a}_p^{(3)}(t)=\quad&\frac{1}{4}\hat{a}_s\hat{B}^\dagger\frac{c_-(\hat{g}_i)}{\hat{N}_i}+\frac{\mathrm{i}}{4}\hat{a}_p\hat{a}_i^\dagger\hat{B}^\dagger\frac{s_-(\hat{g}_i)}{\hat{N}_i^{3/2}}\\
    -&\frac{1}{4}\left(\hat{a}_s\hat{a}_i^2\hat{B}-\hat{a}_p\Delta\hat{N}\hat{N_i}\right)\frac{c_+(\hat{g}_i)}{\hat{N}_i^2}\\
    +&\frac{\mathrm{i}}{4}\left(\hat{a}_s\hat{a}_i\Delta\hat{N}-\hat{a}_p\hat{a}_i\hat{B}\right)\frac{s_+(\hat{g}_i)}{\hat{N}_i^{3/2}}
\end{split}
\end{align}\end{subequations}
where
\begin{subequations}\begin{align}
\begin{split}
    &\hat{B}=\hat{a}_s\hat{a}_p^\dagger,\qquad\Delta\hat{N}=\hat{N}_s-\hat{N}_p,
\end{split}\\
\begin{split}
    &c_\pm(\hat{g}_i)=\cos^3{\hat{g}_i}-\cos{\hat{g}_i}\pm\hat{g}_i\sin{\hat{g}_i},\\
    &s_\pm(\hat{g}_i)=\sin^3\hat{g}_i\pm(\hat{g}_i\cos{\hat{g}_i}-\sin{\hat{g}_i}).
\end{split}
\end{align}\end{subequations}
Taking $g_i\!=\!\langle\hat{N}_i\rangle^{1/2}\chi t\!=\!\pi/2$ and using Eq.~(\ref{coherent average}), we obtain
\begin{subequations}\label{transfer_quantum}\begin{gather}
    \hat{N}_{s,p}(t)\approx\hat{N}_{p,s}\biggl(1-\frac{\pi^2}{16{|\alpha_i|}^2}\biggr),\\
    \operatorname{Var}\hat{N}_{s,p}(t)\approx\operatorname{Var}\hat{N}_{p,s}.
\end{gather}\end{subequations}

Equation~(\ref{transfer_quantum}a) provides an estimation for the photon loss in an ideal quantum-state transfer, and Eq.~(\ref{transfer_quantum}b) indicates that the uncertainty of the transferred quantum field remains unchanged (to third order). This estimation is particularly precise for weak input states of $N_{p,s}\sim1$ or less, where the optimal imaginary gain for the transfer is $g_i=\pi/2$. The probability of photon loss is therefore directly set by the idler intensity. In standard TWM crystals in free space, the idler is typically large (milliwatt level, more than $10^{16}$ photons/s), indicating that the photon loss is negligible. However, if the nonlinearity can be enhanced (e.g., in waveguides, cavities, or other resonant interactions), the needed idler intensity to support the transfer can be much weaker, as low as a few tens of photons \cite{Clark:13, RevModPhys.86.1391, Kockum2019}. Equation~(\ref{transfer_quantum}) provides a rule of thumb to estimate the effect of idler fluctuations on the transfer is such cases.

For stronger inputs with $N_{p,s}\!\gg\!1$, the higher orders contain also a classical contribution due to the average depletion or gain of the strong idler, indicating that we should slightly tune the imaginary gain to preserve the maximal conversion. Since the quantum evolution of Eq.~(\ref{general evolution}) has the same form as the classical equations, we can adjust the gain according to the known classical results \cite{PR1962}. This also indicates that we can determine the classical prediction for the target intensity $N_{s,p}^\mathrm{cl}(t)$ exactly without calculating order by order. Hence, the first correction to the average output intensity is
\begin{subequations}\label{Expectation}
\begin{gather}
\begin{aligned}\begin{split}
   \frac{\bigl\langle\hat{N}_s(t)\bigr\rangle\!-\!N_s^\mathrm{cl}(t)}{\bigl\langle\hat{N}_p\bigr\rangle}&\!\approx \!
    \frac{g_i}{2{|\alpha_i|}^2}\biggl[\frac{1}{2}g_i\cos{2g_i}\!\\
    +\frac{3}{4}\sin{2g_i}+&\left(\!\frac{\operatorname{Var}\hat{N}_p}{\bigl\langle\hat{N}_p\bigr\rangle}\!-\!1\!\right)s_+\!\left(g_i\right)\sin{g_i}\biggr],
\end{split}\end{aligned}\\
\begin{aligned}\begin{split}
   \frac{\bigl\langle\hat{N}_p(t)\bigr\rangle\!-\!N_p^\mathrm{cl}(t)}{\bigl\langle\hat{N}_s\bigr\rangle }&\!\approx\!
    \frac{g_i}{2{|\alpha_i|}^2}\biggl[\frac{1}{2}g_i\cos{2g_i}\!\\
    -\frac{1}{4}\sin{2g_i}-&\left(\!\frac{\operatorname{Var}\hat{N}_s}{\bigl\langle\hat{N}_s\bigr\rangle}\!-\!1\right)s_+\left(g_i\right)\sin{g_i}\biggr].
    \end{split}\end{aligned}
\end{gather}\end{subequations}
With the optimal (imaginary) gain for classical conversion, we can derive the quantum limit on the fidelity of the transfer for a given arbitrary quantum state at the input. The optimal gain for a classical transfer is
\begin{subequations}\begin{align}
    &g_{sp}\!=\!\int_0^1{\frac{\mathrm{d}u}{\sqrt{(1\!-\!u^2)(1\!-\!mu^2)}}}\!\approx\!\frac{\pi}{2}\left(1\!+\!\frac{m}{4}\right),\\
    &g_{ps}=\frac{g_{sp}}{\sqrt{1\!+\!m}}\approx\frac{\pi}{2}\!\left(1-\frac{m}{4}\right),
\end{align}\end{subequations}
where $m\!=\!{\bigl\langle\hat{N}_{x}\bigr\rangle}/{{|\alpha_i|}^2}$ ($x$ represents the input mode to be transferred). Under the approximation $m\!\ll\!1$, Eq.~(\ref{Expectation}) reduces to Eq.~(\ref{transfer_quantum}a) (taking the expectation value), and the variance error of the output intensity will be
\begin{align}\label{Variance}\begin{split}
    \operatorname{Var}\hat{N}_{s,p}&(t)-\operatorname{Var}\hat{N}_{p,s}\!\approx\!\frac{\pi^2m}{16
    }\biggl[1-\frac{\bigl\langle\hat{a}_{p,s}^{\dagger2}\hat{a}_{p,s}^2\bigr\rangle}{{|\alpha_i|}^2}
    \\
    &+\!\frac{2}{{|\alpha_i|}^2}\Bigl(\bigl\langle\hat{N}_{p,s}\hat{a}_{p,s}^{\dagger2}\hat{a}_{p,s}^2\bigr\rangle-\bigl\langle\hat{N}_{p,s}^2\bigr\rangle\bigl\langle\hat{N}_{p,s}\bigr\rangle\Bigr)\!\biggr].
\end{split}\end{align}
The complete expressions for the variance before approximation is in the Appendix.

\begin{figure}
    \centering
    \includegraphics[width=1.0\linewidth]{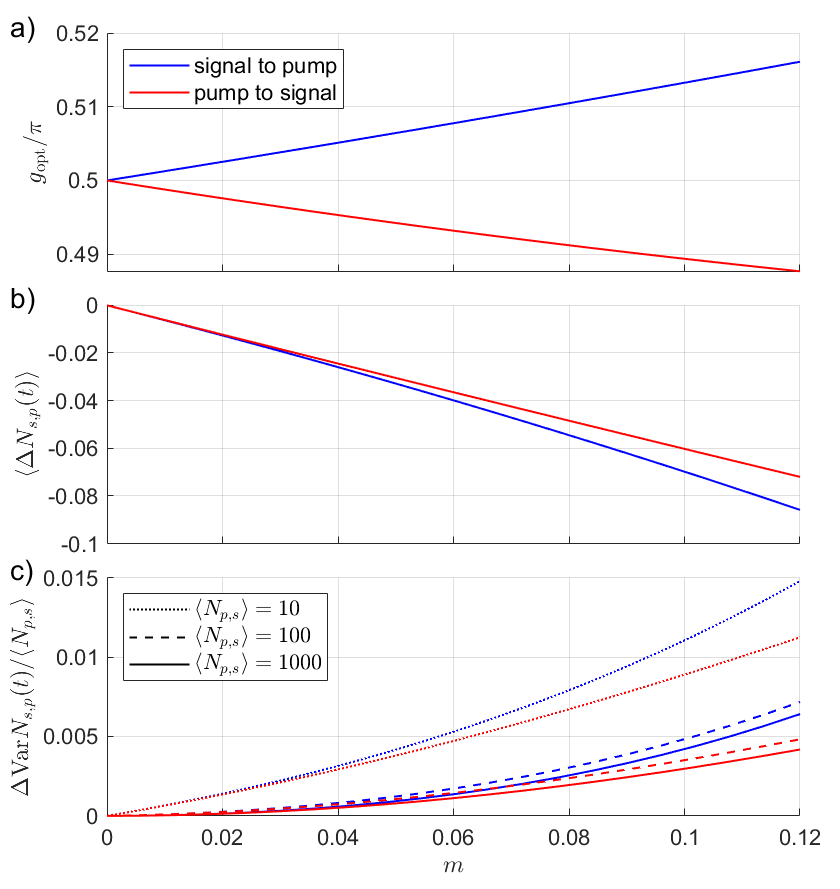}
    \caption{Quantum-state transfer: third-order corrections. (a) Optimal imaginary gain of classical state transfer for varying input intensities. The $x$ axis is $m\!=\!{\bigl\langle\hat{N}_{p,s}\bigr\rangle}/{{|\alpha_i|}^2}$, the photon-number ratio between the input state and the strong idler. (b) Corresponding quantum correction to the average number of transferred photons $\Delta N\!=\!N_{p,s}(t)\!-\!N_{s,p}(0)$. (c) Correction to the variance of the photon number $\Delta\operatorname{Var}{\hat{N}_{s,p}}\!=\!\operatorname{Var}\hat{N}_{s,p}(t)\!-\!\operatorname{Var}\hat{N}_{p,s}(0)$, indicating that the variance of the output depends not only on the $m$ ratio, but also on the absolute intensity of the input state. The blue (red) lines indicate signal-to-pump (pump-to-signal) transfer. The calculations here assumed an input of a weak coherent state.}
    \label{Fig: State transfer}
\end{figure}
Although Eq.~(\ref{Variance}) includes some negative terms, they are all divided by the intensity of the idler, so their contribution in most cases will be very small compared to the first term. Thus, the transfer will introduce an additional uncertainty to the state. The plots for the expectation value and variance of the output state with the exact gain values are shown in Fig.~\ref{Fig: State transfer}. Since the correction highly depends on the state to be transferred, here we set the pump-signal input at the coherent state for convenience; under this assumption, Eq.~(\ref{Variance}) can be simplified to
\begin{equation}\label{var_approx}
    \operatorname{Var}\hat{N}_{s,p}(t)\!-\!\operatorname{Var}\hat{N}_{p,s}\!\approx\!\frac{\pi^2m}{16
    }\left(1+
    m{|\alpha_{p,s}|}^2\right).
\end{equation}
As shown in Fig.~\ref{Fig: State transfer}, the optimal imaginary gain and intensity uncertainty in transfer from the pump to the signal shows a slower dependence on $m$ compared to the signal to the pump, which stems from the small amplification that the down-conversion process can generate to the driving idler for the pump to the signal, but not for the signal to the pump. Since the ratio $m$ is typically small, the correction to the photon number and variance in a single transfer is negligible. Yet, for an integrated device that implements multiple frequency manipulation (during a quantum computation), these corrections can accumulate. Our result enables us to evaluate the overall system performance.
\begin{figure}
    \centering
    \includegraphics[width=1.0\linewidth]{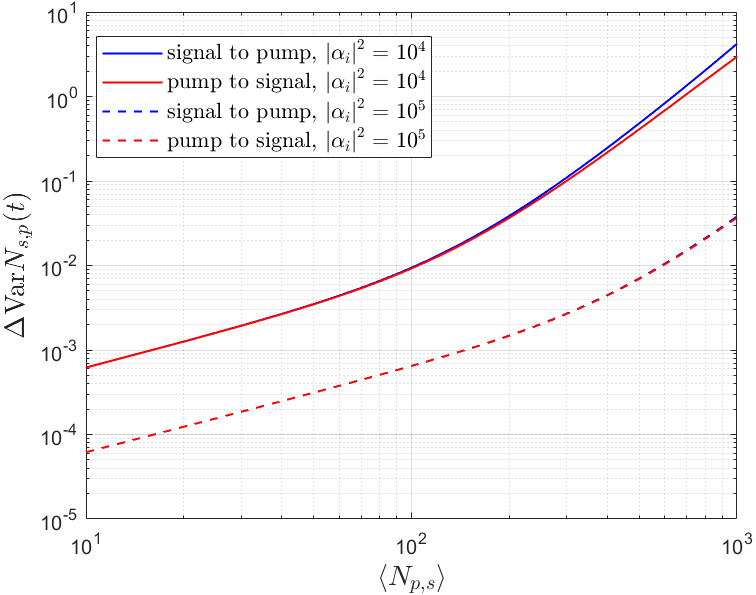}
    \caption{Quantum-state transfer. The quantum correction to the variance of the transferred photon number in the target state is plotted as a function of the input intensity under fixed driving idler: ${|\alpha_i|}^2={10}^4$ (solid lines) and ${|\alpha_i|}^2={10}^5$ (dashed lines). The input state of the pump or signal source is assumed coherent. The two plots for the two cases under ${|\alpha_i|}^2={10}^5$ almost overlap.}
    \label{figure: var with fixed idler}
\end{figure}

Since the intensity of the driving idler is typically fixed, we also plot the variance of the output in the target mode versus the intensity of the source mode (see Fig.~\ref{figure: var with fixed idler}). We find that the variance dependence on the input intensity transforms from linear to cubic at $\bigl\langle\hat{N}_{p,s}\bigr\rangle\!\sim\!{10}^2$, which coincides with Eq.~(\ref{var_approx}) and indicates that the relative fluctuation will increase with the input intensity. However, it is counterintuitive to see increased relative noise under larger transfer intensity. We attribute this result to the neglect of higher-order contributions, which may become important as the $m$ ratio increases, hinting at the validity limits of the expansion to third order.

\section{Conclusion}
In summary, by considering the weak input fields in the TWM processes as a perturbation, we derived a series of time-closed expressions for the time evolution of the coupled fields, which converges extremely fast compared to the standard BCH expansion, indicating its particular utility in the high-gain, highly squeezed regime. Besides the consistency with the existing well-known solutions that assume an undepleted drive [Eqs.~(\ref{previous_transfer}) and (\ref{previous_PDC})], our derivation provides the quantum corrections to the field evolution that considers not only the depletion or generation of the driving field and its quantum fluctuations, but also the commutation relations in the time-differential equations.

By applying our analysis to the PDC and quantum-state transfer processes, we (i) illustrate the coherent measurement of the pump (SFG) as an ideal probe of quantum entanglement by analyzing the dependence of the pump intensity and phase on the input signal or idler fields; (ii) calculate the limit on the squeezing of an ideal OPA with the evolving phase reference of the pump, obtaining a constant limit at large parametric gain instead of the exponential growth of \cite{CrouchBraunstein1988}; and (iii) quantify the lower bound on the fidelity of state transfer due to the quantum corrections to the expectation value and variance of the target-mode output. Although we exemplified our method primarily in the context of quantum optics, the expansion is general and applicable for any trilinear bosonic Hamiltonian, providing high-accuracy analysis to other nonlinear systems as well.

Furthermore, since our method directly provides the time evolution of the field operators with arbitrary inputs, it can easily be expanded to multiprocess Hamiltonians, where linear and nonlinear interactions are cascaded one after the other. A simple example of such a cascaded configuration is the SU(1,1) interferometer where nonlinear interactions are applied sequentially with linear propagation, phase shifts, and loss between them, forming an interferometric sensor whose sensitivity is quantum enhanced by the nonlinear squeezing.

\section*{Acknowledgment}
This research received partial funding from the European QuantERA program (SPARQL project).

\section*{Data availability}
There are no publicly available research data or software supporting this article. Requests for further information or data should be sent to the authors.

\appendix*
\section{Methods}\label{methods}
This Appendix provides further details on the derivations in the paper, including (i) the guiding principles and assumptions of the weak-field perturbative expansion, i.e., how to determine the order of the initial conditions and how to correctly keep or neglect commutators during the iterative solution of each order; (ii) calculation of the expectation values of functions of number operators with coherent-state inputs, which helps to analyze the high-order contribution of strong fields; (iii) detailed derivation up to third order of the field evolution, demonstrated through the example of parametric amplification (the evolution of quantum-state transfer is calculated similarly); and (iv) details of the analysis of quantum-state transfer, i.e., a conceptual explanation of how to deduce the average number of photons at the target mode and its variance from the classical equations and their known solutions.

A note of convention. Throughout this Appendix, canonical operators and number operators that are specified with time are considered as output operators, e.g., $\hat{a}_s(t)$, and unspecified operators are considered as input operators, such as $\hat{a}_s$ and $\hat{N}_p$ standing for $\hat{a}_s(0)$ and $\hat{a}_p^\dagger(0)\hat{a}_p(0)$, respectively.

\subsection{Weak-field expansion: Principles and assumptions}\label{principle details}
We begin with a comprehensive summary of the guiding principles and assumptions of the weak-field expansion. We describe the initial conditions and how their order is determined for the strong and the weak fields and we explain how to treat field commutators during the expansion, i.e., which commutators can be neglected in a given order for either the weak or the strong fields.

\subsubsection{Perturbative assumption and initial conditions}\label{assumption & initial conditions}
We expand the Heisenberg evolution in powers of the weak input field instead of time. Taking parametric amplification, for example, since the process starts with a strong pump, whereas the signal and idler start with a vacuum field, we expand the evolution of each field in powers of $\hat{a}_s^{(\dagger)}$ and $\hat{a}_i^{(\dagger)}$; hence $-\mathrm{i}\hat{a}_s\hat{a}_i\chi t$ is considered a second-order term. We denote by $\hat{a}_{x}^{(k)}(t)$ the $k$th-order contribution to field $x$, or equivalently the sum of all $k$th-order terms in the time evolution of $\hat{a}_{x}(t)$. This expansion converges fast when the change in the strong-field population is small compared to its original value, naturally due to the choice of the small parameter (the population change in the weak field in the interaction equals that in the strong drive).

The initial conditions of the process are also defined according to this principle: Since the pump is strong, its initial condition is $\hat{a}_p^{(0)}(t\!=\!0)\!=\!\hat{a}_p$ for the zeroth order and all higher orders are null with $\hat{a}_p^{(k)}(t\!=\!0)\!=\!0$ for all $k\geq1$. Conversely, for the weak signal or idler, the power $k$ of the operator $\hat{a}^{k}_{s,i}$ reflects the $k$th order of the expansion, indicating that the first order is initialized as $\hat{a}_{s,i}^{(1)}(t\!=\!0)\!=\!\hat{a}_{s,i}$ at the input, and all other orders are null according to $\hat{a}_{s,i}^{(0)}(t\!=\!0)\!=\!0$ and $\hat{a}_{s,i}^{(k)}(t\!=\!0)\!=\!0$ for all $k\geq2$.

\subsubsection{Treatment of commutation relations}\label{comm_order}
The perturbative expansion is generated by an iterative solution of the Heisenberg differential equations (for the time evolution) [Eq.~(\ref{general evolution})]. While on the right-hand side of Eq.~(\ref{general evolution}) the time-dependent operators of different fields commute $\bigl[\hat{a}_s(t),\hat{a}_i(t)\bigr]=\bigl[\hat{a}_{s,i}^\dagger(t),\hat{a}_p(t)\bigr]=0$, the approximate expressions $\hat{a}_x^{(k)}(t)$ do not necessarily commute, and when solving Eq.~(\ref{general evolution}) order by order, we may encounter a situation where the expression to be solved is both left and right multiplied by noncommuting operators [see Eq.~(\ref{s_eq_1st})].

To continue the derivation, we have to reorder the operators and take into account their commutation relations up to the relevant order. To explain how the order of a specific commutator is determined for a given configuration of strong and weak fields, let us take parametric amplification as a guiding example (strong pump and weak signal or idler): The commutator $\bigl[\hat{a}_s,\hat{a}_s^\dagger\bigr]=1$ appears to be of order zero, but it is actually of the same order as the product $\hat{a}_s\hat{a}_s^\dagger$ (second order). Thus, the commutation relation between the weak $\hat{a}_s$ and $\hat{a}_s^\dagger$ must be considered immediately if we want to reorder their product. 

In contrast, the strong pump $\hat{a}_p$ is nearly classical, indicating that the order of $\hat{a}_p\hat{a}^\dagger_p$ or $\hat{a}^\dagger_p\hat{a}_p$ will have a small contribution compared to $\left|\alpha_p\right|^2$. Thus, the commutator $\bigl[\hat{a}_p,\hat{a}_p^\dagger\bigr]\!=\!1$ can be neglected in the zeroth and first orders, but should be included in the second-order solution or higher. Generally, when calculating the $k$th-order term $\hat{a}_x^{(k)}(t)$, we can assume any convenient ordering of the strong-field operators, but when we reach the $(k\!+\!2)$th order, we must take the commutator into account if we wish to reorder the strong-field operators in the term. 

This implies that the form of the $k$th-order correction is not unique and different variant expressions may exist for the same expansion order, depending on how the field operators were arranged in the previous orders of the expansion. However, all the possible variants must be equivalent when writing the total field solution up to the corresponding expansion order. Thus, the derivation of the $k$th-order solution must consider all the previous lower-order solutions, since the $k$th order may require commutation relations that were temporarily neglected previously. For example, the second-order contribution of $\hat{a}_s(t)\hat{a}_i(t)$ is $\hat{a}_s^{(0)}\hat{a}_i^{(2)}\!+\!\hat{a}_s^{(1)}\hat{a}_i^{(1)}\!+\!\hat{a}_s^{(2)}\hat{a}_i^{(0)}$, indicating that when solving the second-order solution, we should also take into account the entire expression up to second order, including the zeroth- and first-order contributions $\hat{a}_s^{(0)}\hat{a}_i^{(0)}\!+\!\hat{a}_s^{(0)}\hat{a}_i^{(1)}\!+\!\hat{a}_s^{(1)}\hat{a}_i^{(0)}$, and consider the (second-order) commutation relations of the pump wherever we wish to exchange the ordering of pump-field operators in the derivation.
\subsection{Mathematical techniques and relation to the classical solutions}
We also introduce some physical and mathematical techniques that help us derive the solutions and calculate the statistical properties of the fields based on these solutions.

\subsubsection{Relation to the classical solution}\label{usage_classical}
Since the form of the Heisenberg differential equations of the quantum evolution is exactly the same as that of the classical evolution, the form of their solutions should also be the same, expect for the importance of the sequential ordering of operators in the quantum expressions. This indicates that some classical conclusions also apply in the quantum case, especially when the quantum commutation relations are negligible or identifiable. For example, we can neglect any commutation relation in the solutions of the zeroth and the first order (according to Appendix~\ref{comm_order} above), so the first-order solution [Eqs.~(\ref{1st order solutions}b), (\ref{State_transfer_1_2}a), and (\ref{State_transfer_1_2}b)] has exactly the same form as the classical solution [Eqs.~(\ref{previous_transfer}) and (\ref{previous_PDC})]. We use this understanding in a less trivial manner in the calculation of the leading corrections to quantum-state transfer (Appendix~\ref{leading correctinos to state transfer}).

\subsubsection{Relation to the BCH formula: Identifying orders}\label{comm_order_id}
As mentioned in Appendix~\ref{comm_order}, the commutation relation of a weak field $[\hat{a}_x,\hat{a}^\dagger_x]=\hat{a}_x\hat{a}^\dagger_x-\hat{a}^\dagger_x\hat{a}_x=1$ is of the same order as the product $\hat{a}_x^\dagger\hat{a}_x$; however, it may also be miscounted as a lower-order term in our derivation since it is a pure number. Fortunately, we can identify whether a term contains any commutators by comparing the power of canonical operators (both strong and weak) with the power of time $t$, because in the Baker-Campbell-Hausdorff expansion (\ref{bch}), the power of time $t$ is always the same as the total order of commutators with the Hamiltonian $\hat{H}$. In the TWM case, every commutation relation with $\bigl[\hat{H}, \ldots\bigr]$ adds one power of canonical operator (either strong or weak) for each term (since each term in $\hat{H}$ is a product of three canonical operators, and the commutation relation reduces the power of operators by 2). Hence, if the power of time is $k$, then the total power of canonical operators should be $k\!-\!1$ assuming no reordering was involved; when we encounter a term where the total power of operators is $k\!-\!1\!-\!2j$, we can infer that it results from reordering operators $j$ times.

\subsubsection{Calculating functional commutation relations}
During the derivation, we encounter many times functional expressions of number operators, such as $\cosh \hat{g}_p\!=\!\cosh{\hat{N}_p^{1/2}\!\chi t}$, so the derivation may include the commutation relation $\bigl[\hat{a}_p,\; \cosh{\hat{N}_p^{1/2}\!\chi t}\bigr]$ between an operator and a function of number operators. Assuming that we expand the function $f\bigl(\hat{N}\bigr)=\sum_m{c_m \hat{N}^m}$, we can calculate these commutators by considering that $\hat{a}\hat{N}^m=\hat{a}\underbrace{\hat{a}^\dagger\hat{a}\hat{a}^\dagger\hat{a}\cdots\hat{a}^\dagger\hat{a}}_{(\hat{a}^\dagger\hat{a})^m}={\bigl(\hat{a}\hat{a}^\dagger\bigr)}^m\hat{a}={\bigl(\hat{N}+1\bigr)}^m\hat{a}$, and similarly $\hat{N}^m\hat{a}^\dagger=\hat{a}^\dagger{\bigl(\hat{N}+1\bigr)}^m$
for any positive integer $m$. Hence, for any power series $f\bigl(\hat{N}\bigr)$,
\begin{equation}\label{func_comm}\begin{aligned}
&\hat{a}f\bigl(\hat{N}\bigr)=f\bigl(\hat{N}+1\bigr)\hat{a},&&f\bigl(\hat{N}\bigr)\hat{a}=\hat{a}f\bigl(\hat{N}-1\bigr),\\
&\hat{a}^\dagger f\bigl(\hat{N}\bigr)=f\bigl(\hat{N}-1\bigr)\hat{a}^\dagger,&&f\bigl(\hat{N}\bigr)\hat{a}^\dagger=\hat{a}^\dagger f\bigl(\hat{N}+1\bigr).
\end{aligned}\end{equation}

\begin{widetext}
\subsection{Quantum corrections for the strong field with coherent-state input}\label{coherent_eval}
Since the guideline of our derivation is to treat the strong field beyond the undepleted approximation, the quantum fluctuations of the strong field should also be taken into account in the high-order analysis of the process. In the absolute majority of practical cases, the state of the strong field at the input can be considered as a coherent state. Thus, we find it useful to explain how the quantum average of functions of number operators can be calculated for coherent states. This is the fundamental calculation for such expressions, not only because the statistical momenta (e.g., variation) of the operator are essentially expectation values of some other functions, but because the expectation value under a coherent state also implies the normal ordering of the expression $\langle\alpha|\!:\!f(\hat{N})\!:\!|\alpha\rangle=f\bigl({|\alpha|}^2\bigr)$, where the normal order of $f\bigl(\hat{N}\bigr)$ is denoted by $:\!f(\hat{N})\!:$ (as a reminder, $:\hat{N}^k:=\hat{a}^{\dagger k}\hat{a}^k$  is the normal order of $\hat{N}^k$).

Let us start with the integer power of a number operator
\begin{equation}\label{expand_N^m}
\hat{N}^m=\sum_{k=0}^m{C(m,k)\!:\!\hat{N^k}\!:},
\end{equation}
where using Eq.~(\ref{func_comm}) we can derive
\begin{equation}
    :\!\hat{N}^k\!:\,=\prod_{j=1}^k{\bigl(\hat{N}-j+1\bigr)},\quad k>0.
\end{equation}
Equation~(\ref{expand_N^m}) corresponds to a known formula about the Stirling numbers of the second kind
\begin{equation}
    x^m=\sum_{k=0}^m{\biggl[S(m,k)\prod_{j=1}^k{(x-j+1)}\biggr]}
\end{equation}
for any $x$ (here we use $x\!=\!\hat{N}$), indicating that the coefficients $C(m,k)$ in Eq.~(\ref{expand_N^m}) are simply the Stirling numbers $\hat{N}^m=\sum_{k=0}^m{S(m,k)\!:\!\hat{N^k}\!:}$.

It is beneficial to rewrite $S(m,k)$ as
\begin{equation}\label{def_R}
    S(m,m-l)=\sum_{k=0}^{m-l}{R(l,k)\binom{m}{k+l}},
\end{equation}
where we take $S(m,m-l)$ as an $(m-l)$th-order polynomial of $m$ and expand it under the bases $\binom{m}{k+l} \;(0\le k\le m-l)$ with $R(l,k)$ the coefficients, since this allows us to represent $\langle\alpha|f(\hat{N})|\alpha\rangle$ as
\begin{equation}\begin{aligned}\label{full quantum average}
\langle\alpha|f(\hat{N})|\alpha\rangle=&\sum_m{\frac{\langle\alpha|\hat{N}^m|\alpha\rangle}{m!}}=\sum_{m,l}{\frac{S(m,m-l)}{m!}{|\alpha|}^{2(m-l)}}\\
=&\sum_{m,k,l}{\frac{R(l,k){|\alpha|}^{2(m-l)}}{(m-k-l)!(k+l)!}}=\sum_{k,l}{\frac{R(l,k)}{(k+l)!}{|\alpha|}^{2k}f^{(k+l)}{\bigl({|\alpha|}^2\bigr)}}.
\end{aligned}\end{equation}
Plugging Eq.~(\ref{def_R}) into the recurrence formula for the Stirling numbers of the second kind $S(m,k)=k\cdot S(m-1,k)+S(m-1,k-1)$, $S(0,0)=1$, and $S(0,k)=0\,(k\neq0)$, we can obtain
\begin{equation}\label{direct_R}
    \sum_{k=0}^{m-l}{R(l,k)\binom{m}{k+l}}=(m-l)\sum_{k=0}^{m-l}{R(l-1,k)\binom{m-1}{k+l-1}}+\sum_{k=0}^{m-l-1}{R(l,k)\binom{m-1}{k+l}}.
\end{equation}

We can derive the recurrence of $R(l,k)$ by considering
\begin{equation}\begin{aligned}\label{R_deriv}
    \sum_{k=0}^{m-l}{R(l,k)\binom{m-1}{k+l-1}}=&\sum_{k=0}^{m-l}{R(l,k)\binom{m}{k+l}}-\sum_{k=0}^{m-l}{R(l,k)\binom{m-1}{k+l}}\\
    =&\sum_{k=0}^{m-l}{[(m-l-k)+k]R(l-1,k)\binom{m-1}{k+l-1}}\\
    =&\sum_{k=0}^{m-l}{R(l-1,k)\left[(k+l)\binom{m-1}{k+l}+k\binom{m-1}{k+l-1}\right]}\\
    =&\sum_{k=1}^{m-l}{(k+l-1)R(l-1,k-1)\binom{m-1}{k+l-1}}+\sum_{k=0}^{m-l}{kR(l-1,k)\binom{m-1}{k+l-1}},
\end{aligned}\end{equation}
where the first and third equalities utilize the properties of combination numbers and the second applies Eq.~(\ref{direct_R}). Hence

\begin{equation}\begin{aligned}\label{R_recur}
    &R(0,0)=1,\quad R(0,k)=0\,(k\ne0),
    &R(l,k)=k\!\cdot\!R(l\!-\!1,k)+(k\!+\!l\!-\!1)\!\cdot\!R(l\!-\!1,k\!-\!1).
\end{aligned}\end{equation}

The flipped triangle $R(l,l-k+1)$ is partially listed in the convolution matrix of the double factorial of odd numbers in \cite{OEIS_A035342}. We prefer the flipped form [the definition of Eq.~(\ref{def_R})] because Eq.~(\ref{full quantum average}) is expanded directly in powers of $1/{|\alpha|^2}$ with $l$ as the exponent of each term (note that each order of derivative with respect to $\hat{N}$ around ${|\alpha|}^2$ also reduces the power of $|\alpha|$ by $2$), which indicates that we can obtain the $2l_0$th-order approximation for $\bigl\langle f\bigl(\hat{N}\bigr)\bigr\rangle$ by cutting the summation in Eq.~(\ref{full quantum average}) at $l=l_0$. Hence, the classical approximation $\bigl\langle f\bigl(\hat{N}\bigr)\bigr\rangle\!\approx\!f\bigl({|\alpha|}^2\bigr)$ that is used in the undepleted analysis only considers the term with $R(0,0)$ in Eq.~(\ref{full quantum average}); we can identify its first nonvanishing correction by considering the only nonzero term with $l=1$, which is $\frac{1}{2}{|\alpha|}^2f''\bigl({|\alpha|}^2\bigr)$ taken at $k=1$ and leads to Eq.~(\ref{coherent average}) immediately. This can be continued up to any desired order, and the quantum uncertainty (variance) can also be calculated by considering the squared function. 

\subsection{Detailed derivation up to third order (parametric amplification)}\label{derivation_3rd}
Let us provide a detailed derivation example for the weak-field expansion up to third order through the process of parametric amplification (the derivation for quantum-state transfer follows the exact same procedure).

Classically, when a monochromatic pump enters a nonlinear medium, no photon will be down-converted without signal or idler seeding. Under our principles of analysis, this forms our zeroth-order solution of the process $\hat{a}_p^{(0)}(t)=\hat{a}_p$, $\hat{a}_{s,i}^{(0)}(t)\equiv0$; this solution can also be obtained directly by considering the initial conditions mentioned in Appendix~\ref{assumption & initial conditions}.

In the derivation of the first-order solution, we have
\begin{equation}
    \frac{\mathrm{d}}{\mathrm{d}t}\hat{a}_{s,i}^{(1)}(t)=-\mathrm{i}\chi\hat{a}_{i,s}^{(1)\dagger}(t)\hat{a}_p^{(0)}(t),
\end{equation}
and taking the second derivative of the equation yields
\begin{equation}\label{s_eq_1st}
    \frac{\mathrm{d}^2}{{\mathrm{d}t}^2}\hat{a}_s^{(1)}(t)=\hat{a}_p^\dagger\hat{a}_s^{(1)}(t)\hat{a}_p.
\end{equation}
Although the expression of $\hat{a}_s^{(1)}(t)$ may include $\hat{a}_p^{(\dagger)}$, the commutation relation in the pump field only contribute to the third or higher order (adding at least two orders to the first) and hence can be neglected under the first-order approximation; so $\hat{a}_s^{(1)}(t)$ and $\hat{a}_p$ can be treated as if they commute, and the solution can be written as
\begin{equation}\label{sol_1st}
    \hat{a}_{s,i}^{(1)}(t)=\hat{a}_{s,i}\cosh{\hat{g}_p}-\frac{\mathrm{i}\hat{a}_{i,s}^\dagger\hat{a}_p}{\hat{N}_p^{1/2}}\sinh{\hat{g}_p},
\end{equation}
where $\hat{g}_p=\hat{N}_p^{1/2}\chi t$.

To calculate the second order (for the pump field only since the second-order solution for the signal and idler is null), we can directly plug the first-order solutions into the second-order perturbative equation (we consider the multimode case here)
\begin{align}\begin{split}\label{eqn_2nd}
    \frac{\mathrm{d}}{\mathrm{d}t}\hat{a}_p^{(2)}(t)=&-\mathrm{i}\chi\sum_{\omega_s+\omega_i=\omega_p}{\hat{a}_s^{(1)}(t)\hat{a}_i^{(1)}(t)}\\
    =&-\mathrm{i}\chi\sum_{\omega_s+\omega_i=\omega_p}{\left(\hat{a}_s\cosh{\hat{g}_p}-\frac{\mathrm{i}\hat{a}_i^\dagger\hat{a}_p}{\hat{N}_p^{1/2}}\sinh{\hat{g}_p}\right)\left(\hat{a}_i\cosh{\hat{g}_p}-\frac{\mathrm{i}\hat{a}_s^\dagger\hat{a}_p}{\hat{N}_p^{1/2}}\sinh{\hat{g}_p}\right)}.
\end{split}\end{align}
Here again the commutation relation between the pump operators can be neglected, so Eq.~(\ref{eqn_2nd}) can be integrated as
\begin{equation}\label{sol_2nd}
    \hat{a}_p^{(2)}(t)=-\frac{\mathrm{i}\hat{A}}{2\hat{N}_p^{1/2}}\sinh{2\hat{g}_p}-\hat{a}_p\frac{\hat{N}_\mathrm{dc}}{2\hat{N}_p}\sinh^2{\hat{g}_p}+\frac{\mathrm{i}\hat{H}'\hat{a}_p}{4\hat{N}_p^{3/2}}(\sinh{2\hat{g}_p}-2\hat{g}_p),
\end{equation}
where $\hat{A}$ and $\hat{N}_\mathrm{dc}$ follow the definition of Eq.~(\ref{A & N_dc}) and $\hat{H}'=\hat{A}\hat{a}_p^\dagger+\hat{A}^\dagger\hat{a}_p$ is the unitless multimode Hamiltonian (multiple signal-idler pairs). The $+1$ in $\hat{N}_\mathrm{dc}$ arises from the commutation relation of the signal operators and cannot be neglected. The same procedure can also be applied to the case of quantum-state transfer with a strong idler and a weak signal and pump [Eq.~(\ref{State_transfer_1_2})].

The third-order solution can be derived by keeping Eq.~(\ref{general evolution}) up to the third order
\begin{equation}\label{eqn_3rd_expanded}
    \frac{\mathrm{d}}{\mathrm{d}t}\Bigl[\hat{a}_{s,i}^{(1)}(t)+\hat{a}_{s,i}^{(3)}(t)\Bigr]=-\mathrm{i}\chi\Bigl[\hat{a}_{i,s}^{(1)\dagger}(t)\hat{a}_p^{(0)}(t)+\hat{a}_{i,s}^{(1)\dagger}(t)\hat{a}_p^{(2)}(t)+\hat{a}_{i,s}^{(3)\dagger}(t)\hat{a}_p^{(0)}(t)\Bigr].
\end{equation}
Taking the second derivative of Eq.~(\ref{eqn_3rd_expanded}) yields the following equation for $\hat{a}_s^{(3)}(t)$: 
\begin{equation}\label{s_eq_3rd}\begin{aligned}
    \frac{\mathrm{d}^2}{{\mathrm{d}t}^2}\hat{a}_s^{(3)}(t)=&-\mathrm{i}\chi\left[\frac{\mathrm{d}}{\mathrm{d}t}\hat{a}_i^{(1)\dagger}(t)\hat{a}_p^{(0)}(t)+\frac{\mathrm{d}}{\mathrm{d}t}\hat{a}_i^{(1)}(t)\hat{a}_p^{(2)}(t)+\hat{a}_i^{(1)}(t)\frac{\mathrm{d}}{\mathrm{d}t}\hat{a}_p^{(2)}(t)\right]\\
    &+\chi^2\Bigl[\hat{a}_p^{(2)\dagger}(t)\hat{a}_s^{(1)}(t)+\hat{a}_p^{(0)\dagger}(t)\hat{a}_s^{(3)}(t)\Bigr]\hat{a}_p^{(0)}(t)-\frac{\mathrm{d}^2}{{\mathrm{d}t}^2}\hat{a}_s^{(1)}(t)\\
    =&\quad\,\chi^2\Bigl[\hat{a}_p^{(0)\dagger}(t)\hat{a}_s^{(1)}(t)\hat{a}_p^{(2)}(t)+\hat{a}_p^{(2)\dagger}(t)\hat{a}_s^{(1)}(t)\hat{a}_p^{(0)}(t)-\hat{a}_i^{(1)\dagger}(t)\hat{a}_s^{(1)}(t)\hat{a}_i^{(1)}(t)\Bigr]\\
    &+\underline{\chi^2\hat{a}_p^{(0)\dagger}(t)\hat{a}_s^{(3)}(t)\hat{a}_p^{(0)}(t)}+\left[\chi^2\hat{a}_p^{(0)\dagger}(t)\hat{a}_s^{(1)}(t)\hat{a}_p^{(0)}(t)-\frac{\mathrm{d}^2}{{\mathrm{d}t}^2}\hat{a}_s^{(1)}(t)\right]\\
    \approx&\quad\,\chi^2\left\{\left[\hat{a}_p^{(0)\dagger}(t)\hat{a}_p^{(2)}(t)+\hat{a}_p^{(2)\dagger}(t)\hat{a}_p^{(0)}(t)\right]\hat{a}_s^{(1)}(t)+\hat{a}_p^{(0)}(t)\left[\hat{a}_s^{(1)}(t),\hat{a}_p^{(2)}(t)\right]-\hat{a}_i^{(1)\dagger}(t)\hat{a}_s^{(1)}(t)\hat{a}_i^{(1)}(t)\right\}\\
    &+\underline{\chi^2\hat{N}_p\hat{a}_s^{(3)}(t)}+\left[\chi^2\hat{a}_p^{(0)\dagger}(t)\hat{a}_s^{(1)}(t)\hat{a}_p^{(0)}(t)-\frac{\mathrm{d}^2}{{\mathrm{d}t}^2}\hat{a}_s^{(1)}(t)\right],
\end{aligned}\end{equation}
Here the final equivalence only comes from exchanging $\hat{a}_s^{(3)}(t)$ and $\hat{a}_p$ in the underlined term, which will affect only the fifth order or higher.
Let us now calculate the terms on the right-hand side of Eq.~(\ref{s_eq_3rd}) one by one, taking Eq.~(\ref{sol_1st}) as $\hat{a}_{s,i}^{(1)}(t)$, the first-order solution for the signal or idler, and Eq.~(\ref{sol_2nd}) as $\hat{a}_{p}^{(2)}(t)$, the second-order solution for the pump. We start with $\hat{a}_p^{(0)\dagger}(t)\hat{a}_s^{(1)}(t)\hat{a}_p^{(0)}(t)$,
\begin{equation}\label{comm_1}\begin{aligned}
    \hat{a}_p^{(0)\dagger}(t)\hat{a}_s^{(1)}(t)\hat{a}_p^{(0)}(t)=\,&\hat{a}_p^\dagger\left(\hat{a}_s\cosh{\hat{g}_p}-\frac{\mathrm{i}\hat{a}_i^\dagger\hat{a}_p}{\hat{N}_p^{1/2}}\sinh{\hat{g}_p}\right)\hat{a}_p=\hat{a}_s\hat{a}_p^\dagger\cosh{\hat{g}_p}\;\hat{a}_p-\mathrm{i}\hat{a}_i^\dagger\hat{N}_p^{1/2}\sinh{\hat{N}_p^{1/2}\chi t}\;\hat{a}_p\\
    =\,&\hat{a}_s\hat{N}_p\cosh{{\bigl(\hat{N}_p-1\bigr)}^{1/2}\chi t}-\mathrm{i}\hat{a}_i^\dagger\hat{a}_p{\bigl(\hat{N}_p-1\bigr)}^{1/2}\sinh{{\bigl(\hat{N}_p-1\bigr)}^{1/2}\chi t}\\
    \approx\,&\hat{N}_p\hat{a}_s^{(1)}(t)-\frac{1}{2}\left[\hat{a}_s\hat{g}_p\sinh{\hat{g}_p}-\frac{\mathrm{i}\hat{a}_i^\dagger\hat{a}_p}{\hat{N}_p^{1/2}}\bigl(\hat{g}_p\cosh{\hat{g}_p}+\sinh{\hat{g}_p}\bigr)\right]
\end{aligned}\end{equation}
where Eq.~(\ref{func_comm}) was applied in the second line and the equivalence in the last line represents a Taylor approximation, since $1$ is much smaller than $\hat{N}_p$. Hence
\begin{equation}\label{from_1st}
    \chi^2\hat{a}_p^{(0)\dagger}(t)\hat{a}_s^{(1)}(t)\hat{a}_p^{(0)}(t)-\frac{\mathrm{d}^2}{{\mathrm{d}t}^2}\hat{a}_s^{(1)}(t)=-\frac{\chi^2}{2}\left[\hat{a}_s\hat{g}_p\sinh{\hat{g}_p}-\frac{\mathrm{i}\hat{a}_i^\dagger\hat{a}_p}{\hat{N}_p^{1/2}}\bigl(\hat{g}_p\cosh{\hat{g}_p}+\sinh{\hat{g}_p}\bigr)\right].
\end{equation}
We then calculate the expressions
\begin{subequations}\label{more_terms}\begin{align}
\begin{split}
    &\hat{a}_p^{(0)\dagger}(t)\hat{a}_p^{(2)}(t)+\hat{a}_p^{(2)\dagger}(t)\hat{a}_p^{(0)}(t)=-\hat{N}_\mathrm{dc}\sinh^2{\hat{g}_p}+\left(\hat{A}^\dagger\hat{a}_p-\hat{A}\hat{a}_p^\dagger\right)\frac{\mathrm{i}}{\hat{N}_p^{1/2}}\cosh{\hat{g}_p}\sinh{\hat{g}_p},
\end{split}\\
\begin{split}
    &\hat{a}_p^{(0)}(t)\left[\hat{a}_s^{(1)},\hat{a}_p^{(2)}\right]=\frac{\hat{a}_s}{2}\hat{g}_p\sinh{\hat{g}_p}+\frac{\mathrm{i}\hat{a}_i^\dagger\hat{a}_p}{2\hat{N}_p^{1/2}}\bigl(\sinh{\hat{g}_p}-\hat{g}_p\cosh{\hat{g}_p}\bigr),
\end{split}\\
\begin{split}
    &\hat{a}_i^{(1)\dagger}(t)\hat{a}_s^{(1)}(t)\hat{a}_i^{(1)}(t)=\quad\hat{a}_i^\dagger\cosh{\hat{g}_p}\left(\hat{A}\cosh^2{\hat{g}_p}-\frac{\mathrm{i}\hat{N}_\mathrm{dc}\hat{a}_p}{\hat{N}_p^{1/2}}\cosh{\hat{g}_p}\sinh{\hat{g}_p}-\frac{\hat{a}_p^2\hat{A}^\dagger}{\hat{N}_p}\sinh^2{\hat{g}_p}\right)\\
    &\qquad\qquad\qquad\qquad\qquad+\hat{a}_s\sinh{\hat{g}_p}\left(\frac{\mathrm{i}\hat{a}_p^\dagger\hat{A}}{\hat{N}_p^{1/2}}\cosh^2{\hat{g}_p}+\hat{N}_\mathrm{dc}\cosh{\hat{g}_p}\sinh{\hat{g}_p}-\frac{\mathrm{i}\hat{a}_p\hat{A}^\dagger}{\hat{N}_p^{1/2}}\sinh^2{\hat{g}_p}\right)
\end{split}
\end{align}\end{subequations}
where we consider again the multimode case with the notation of Eq.~(\ref{A & N_dc}). Plugging Eqs.~(\ref{from_1st}) and (\ref{more_terms}) into Eq.~(\ref{s_eq_3rd}), we obtain
\begin{equation}\label{s_eq_3rd_f}\begin{aligned}
    \frac{\mathrm{d}^2}{{\mathrm{d}t}^2}\hat{a}_s^{(3)}(t)=&-\hat{a}_i^\dagger\hat{A}\cosh{2\hat{g}_p}\cosh{\hat{g}_p}+\frac{\mathrm{i}\hat{a}_i^\dagger\hat{a}_p\hat{N}_\mathrm{dc}}{\hat{N}_p^{1/2}}\cosh{2\hat{g}_p}\sinh{\hat{g}_p}+\frac{\hat{a}_i^\dagger\hat{a}_p^2\hat{A}^\dagger}{\hat{N}_p}\sinh{2\hat{g}_p}\sinh{\hat{g}_p}\\
    &-\frac{\mathrm{i}\hat{a}_s\hat{a}_p^\dagger\hat{A}}{\hat{N}_p^{1/2}}\cosh{\hat{g}_p}\sinh{2\hat{g}_p}-\hat{a}_s\hat{N}_\mathrm{dc}\sinh{2\hat{g}_p}\sinh{\hat{g}_p}+\frac{\mathrm{i}\hat{a}_s\hat{a}_p\hat{A}^\dagger}{\hat{N}_p^{1/2}}\cosh{2\hat{g}_p}\sinh{\hat{g}_p}.
\end{aligned}\end{equation}

To solve Eq.~(\ref{s_eq_3rd_f}), we rewrite the products of the hyperbolic functions as sums
\begin{equation}\begin{aligned}
    &\cosh{2x}\cosh{x}=\frac{1}{2}\left(\cosh{3x}+\cosh{x}\right),&&\sinh{2x}\sinh{x}=\frac{1}{2}\left(\cosh{3x}-\cosh{x}\right),\\
    &\cosh{2x}\sinh{x}=\frac{1}{2}\left(\sinh{3x}-\sinh{x}\right),&&\cosh{x}\sinh{2x}=\frac{1}{2}\left(\sinh{3x}+\sinh{x}\right),
\end{aligned}\end{equation}
and use the special solution to the nonhomogeneous linear differential equation $y''(x)=k^2y(x)+P_3\cosh{3kx}+Q_3\sinh{3kx}+P_1\cosh{kx}+Q_1\sinh{kx}$
with initial conditions $y(0)=0$ and $y'(0)=0$ as
\begin{equation}
y(x)=\frac{P_3}{2k^2}(\cosh^3{kx}-\cosh{kx})+\frac{Q_3}{2k^2}\sinh^3{kx}+\frac{P_1}{2k}x\sinh{kx}+\frac{Q_1}{2k^2}(kx\cosh{kx}-\sinh{kx}),
\end{equation}
which yields the third-order solution (\ref{sol_3rd}). 

\textit{Sanity check by comparison to the BCH expansion}. The solutions (\ref{sol_1st}) and (\ref{sol_3rd}) can be verified by comparing their power expansion in time with the BCH expansion, which essentially rearranges the same terms in a different way. Specifically, we can implement a sanity check for the current results by calculating all the third-order terms in the BCH formula and some relevant fourth-order terms
\begin{align}
\begin{split}
    \hat{a}_s(t)=\quad&\hat{a}_s-\mathrm{i}(\chi t)\hat{a}_i^\dagger\hat{a}_p+\frac{{(\chi t)}^2}{2}\bigl(\hat{a}_s\hat{N}_p-\hat{a}_i^\dagger\hat{A}\bigr)-\frac{\mathrm{i}{(\chi t)}^3}{6}\bigl(\hat{a}_i^\dagger\hat{a}_p\hat{N}_p+2\hat{a}_s\hat{A}\hat{a}_p^\dagger-\hat{a}_s\hat{A}^\dagger\hat{a}_p-\hat{a}_i^\dagger\hat{N}_\mathrm{dc}\hat{a}_p\bigr)\\
    +&\frac{{(\chi t)}^4}{24}\bigl(\hat{a}_s \hat{N}_p^2-4\hat{a}_s\hat{N}_\mathrm{dc}\hat{N}_p-6\hat{a}_i^\dagger\hat{A}\hat{N}_p+4\hat{a}_i^\dagger\hat{a}_p^2\hat{A}^\dagger\bigr)+\cdots,
\end{split}\end{align}
and compare them with the Taylor approximation of Eq.~(\ref{CpmSpm}),
\begin{align}
    &C_+(x)\approx 2x^2+x^4,
    &C_-(x)\approx 2x^4/3,
    &&S_+(x)\approx 4x^3/3,
    &&&S_-(x)\approx 2x^3/3.
\end{align}

For the state transfer process, we can follow the same sanity check
\begin{align}
\begin{split}
    \hat{a}_s(t)=\quad&\hat{a}_s-\mathrm{i}(\chi t)\hat{a}_i^\dagger\hat{a}_p-\frac{{(\chi t)}^2}{2}\bigl(\hat{a}_s\hat{N}_i-\hat{B}\hat{a}_p\bigr)-\frac{\mathrm{i}{(\chi t)}^3}{6}\bigl(2\hat{a}_s\hat{a}_i\hat{B}-\hat{a}_s\hat{a}_i^\dagger\hat{B}^\dagger-\hat{a}_p\hat{N}_i\hat{a}_i^\dagger-\hat{a}_p\hat{a}_i^\dagger\Delta\hat{N}\bigr)\\
    +&\frac{{(\chi t)}^4}{24}\bigl(\hat{a}_s\hat{N}_i^2+4\hat{a}_p\hat{a}_i^{\dagger2}\hat{B}^\dagger-6\hat{a}_p\hat{B}\hat{N}_i+4\Delta\hat{N}\hat{a}_s\hat{N}_i\bigr)+\cdots
\end{split}\\
\begin{split}
    \hat{a}_p(t)=\quad&\hat{a}_p-\mathrm{i}(\chi t)\hat{a}_s\hat{a}_i-\frac{{(\chi t)}^2}{2}\bigl(\hat{a}_s\hat{B}^\dagger+\hat{a}_p\hat{N}_i\bigr)-\frac{\mathrm{i}{(\chi t)}^3}{6}\bigl(\hat{a}_p\hat{a}_i\hat{B}-\hat{a}_s\hat{a}_i\hat{N}_i-2\hat{a}_p\hat{a}_i^\dagger\hat{B}^\dagger-\hat{a}_s\hat{a}_i\Delta\hat{N}\bigr)\\
    +&\frac{{(\chi t)}^4}{24}\bigl(\hat{a}_p\hat{N}_i^2-4\hat{a}_s\hat{a}_i^2\hat{B}+6\hat{a}_s\hat{B}^\dagger\hat{N}_i+4\hat{a}_p\Delta\hat{N}\hat{N}_i\bigr)+\cdots
\end{split}
\end{align}
and
\begin{equation}\begin{aligned}
    &c_+(x)=\cos^3{x}-\cos{x}+x\sin{x}=2x^4/3+\cdots,
    &&c_-(x)=\cos^3{x}-\cos{x}-x\sin{x}=-2x^2+x^4+\cdots,\\
    &s_+(x)=\sin^3{x}+(x\cos{x}-\sin{x})=2x^3/3+\cdots,
    &&s_-(x)=\sin^3{x}-(x\cos{x}-\sin{x})=4x^3/3+\cdots,
\end{aligned}\end{equation}
which verifies the coefficients for the third-order solutions (\ref{State_transfer_3}).

\subsection{Quantum-state transfer: Corrections to the target-state population (expectation and variance)}\label{leading correctinos to state transfer}

As we have obtained the expressions of the field evolution up to third order in Eqs.~(\ref{State_transfer_1_2}a) and (\ref{State_transfer_3}a), we can calculate the expectation value and variance of the photon number in each field (to third order). We start by considering the time-dependent operator for the photon number at the target state (we take pump-to-signal transfer as an example):
\begin{equation}\label{expectation_4}
\hat{N}_s(t)=\hat{a}_s^\dagger(t)\hat{a}_s(t) \approx\hat{a}_s^{(1)\dagger}(t)\hat{a}_s^{(1)}(t)+\hat{a}_s^{(1)\dagger}(t)\hat{a}_s^{(3)}(t)+\hat{a}_s^{(3)\dagger}(t)\hat{a}_s^{(1)}(t).
\end{equation}
The time-dependent variance of the photon number is then
\begin{equation}\label{variance_4}
    \operatorname{Var}{\hat{N}_s(t)}\approx\Bigl\langle{\bigl[\hat{a}_s^{(1)\dagger}(t)\hat{a}_s^{(1)}(t)\bigr]}^2+\bigl[\hat{a}_s^{(3)\dagger}(t)\hat{a}_s^{(1)}(t)\hat{a}_s^{(1)\dagger}(t)\hat{a}_s^{(1)}(t)+\hat{a}_s^{(1)\dagger}(t)\hat{a}_s^{(1)}(t)\hat{a}_s^{(3)\dagger}(t)\hat{a}_s^{(1)}(t)+c.c.\bigr]\Bigr\rangle-{\bigl\langle\hat{N}_s(t)\bigr\rangle}^2.
\end{equation}
Let us first consider the field operators at the time of optimal transfer $t_{\pi/2}$. In the solution of the correction to the signal target field $\hat{a}_s^{(3)}(t_{\pi/2})$, we can again take the strong idler as purely classical and substitute $\hat{g}_i\!=\!\pi/2$ (nontrivially, this is still the optimal gain for the transfer up to the third order), since any quantum correction or small classical depletion to the strong idler will contribute only to the fifth order or higher. Thus, $\hat{a}_s^{(3)}(t)$ simplifies to
\begin{equation}\label{3rd_half_pi}
     \hat{a}_s^{(3)}(t_{\pi/2})=\mathrm{i}\frac{\hat{a}_s\hat{a}_i\hat{N}_p}{{|\alpha_i|}^{3}}.
\end{equation}

When calculating $\bigl\langle\hat{N}_s(t_{\pi/2})\bigr\rangle$ and $\operatorname{Var}{\hat{N}_s(t_{\pi/2})}$ the contribution of Eq.~(\ref{3rd_half_pi}) cancels out due to its pure-imaginary coefficient and the vacuum input of the signal state (which partially explains why $\hat{g}_i\!=\!\pi/2$ remains optimal without correction up to the third order). However, the quantum fluctuations of the driving field do contribute to the average photon number of the target according to $\bigl\langle\hat{N}_s(t_{\pi/2})\bigr\rangle\!=\!\bigl\langle\hat{N}_p(0)\bigr\rangle\bigl(1-\pi^2/16{|\alpha_i|}^2\bigr)$ [as stated in Eq.~(\ref{transfer_quantum}a)], which was calculated using Eq.~(\ref{coherent average}a).

We note that for state transfer, the optimal gain depends slightly on the average population $\langle\hat{N}_p\rangle$ to be transferred from the source to the target state. When $\langle\hat{N}_p\rangle\!\sim\!1$, the source and target states must be handled fully quantum mechanically, which mandates $\hat{g}_i\!=\!\pi/2$ to third order (corrections may appear in higher orders). However, for a larger transferred population ($1\!\ll\!\langle\hat{N}_p\rangle\!\ll\!{|\alpha_i|}^2$), it is possible to treat the source and target semiclassically and keep the commutation relations only up to the third order. In that case, the value of $\hat{g}_i$ can be further optimized by considering the exact classical solution, taking into account the small depletion of the driving idler. This is justified because the classical solution is equivalent to the quantum solution after all the commutation relations are omitted from the final expression (see Appendix~\ref{usage_classical}), and the contribution of commutation relations can be easily identified in any expression using the technique in Appendix~\ref{comm_order_id} (comparing the exponent of time and canonical operators). Hence, it is safe to cut off the contribution of the commutation relations at some certain order with respect to the (semiclassical) source field $\hat{a}_p$. Although the exact classical solution for state transfer has been fully derived in the literature \cite{PR1962}, our derivation does not require most of its details or its precise expression; we only need the optimal classical gain, where the target pump at the output reproduces exactly the source signal at the input.

When the signal input is vacuum, only one third-order term in Eq.~(\ref{expectation_4}) contributes to the expectation value of $\bigl\langle\hat{N}_s(t)\bigr\rangle$,
\begin{equation}
    \bigl\langle\hat{a}_s^{(1)\dagger}(t)\hat{a}_s^{(3)}(t)+c.c.\bigr\rangle=\frac{\langle\hat{a}_p^{\dagger2}\hat{a}_p^2\rangle}{2{|\alpha_i|}^2}\sin{g_i}s_+(g_i),
\end{equation}
whose classical counterpart is 
\begin{equation}
    {\bigl\langle\hat{a}_s^{(1)\dagger}(t)\hat{a}_s^{(3)}(t)+c.c.\bigr\rangle}_{cl}=\frac{{\bigl\langle\hat{N}_p\bigr\rangle}^2}{2{|\alpha_i|}^2}\sin{g_i}s_+(g_i).
\end{equation}
Thus, we derived the quantum correction to the transferred population due to the pump source state [the last term in Eq.~(\ref{Expectation}a)], and the other terms arise from the quantum fluctuation of the strong idler drive, as discussed previously.

The calculation of the variances $\operatorname{Var}\hat{N}_{s,p}(t_\mathrm{opt})$ is slightly more complicated because more terms in the expression of $\hat{a}_x^{(3)}(t)$ should be taken into account. At the optimal transfer gain, we obtain 
\begin{subequations}
\begin{gather}
\begin{aligned}\begin{split}
&\operatorname{Var}\hat{N}_s(t_\mathrm{opt})-\operatorname{Var}\hat{N}_p\!\approx 2\left\langle\bigl(\hat{N}_p\sin^2{g_{ps}}-\bigl\langle\hat{N}_p\bigr\rangle\bigr)\bigl(\hat{N}_s(t_\mathrm{opt})-\hat{N}_p\bigr)\right\rangle+\frac{1}{4}\bigl\langle\hat{N}_s(2t_\mathrm{opt})\bigr\rangle+\frac{\bigl\langle\hat{N}_p^2\bigr\rangle}{4{|\alpha_i|}^2}g_{ps}^2\sin^2{2g_{ps}}\\
&-\frac{\langle\hat{a}_p^{\dagger2}\hat{a}_p^2\rangle}{2{|\alpha_i|^2}}\!\sin^3{g_{ps}}s_+(g_{ps})+\frac{\operatorname{Var}\hat{N}_p-\bigl\langle\hat{N}_p\bigr\rangle}{4{|\alpha_i|}^2}\sin^3{g_{ps}}\cos{g_{ps}}\left(\sin{4g_{ps}}-\sin{2g_{ps}}-2g_{ps}\right),
\end{split}\end{aligned}\\
\begin{aligned}\begin{split}
&\operatorname{Var}\hat{N}_p(t_\mathrm{opt})-\operatorname{Var}\hat{N}_s\!\approx 2\left\langle\bigl(\hat{N}_s\sin^2{g_{sp}}-\bigl\langle\hat{N}_s\bigr\rangle\bigr)\bigl(\hat{N}_p(t_\mathrm{opt})-\hat{N}_s\bigr)\right\rangle+\frac{1}{4}\bigl\langle\hat{N}_p(2t_\mathrm{opt})\bigr\rangle+\frac{\bigl\langle\hat{N}_s^2\bigr\rangle}{4{|\alpha_i|}^2}g_{sp}^2\sin^2{2g_{sp}}\\
&+\frac{\sin^3{g_{sp}}}{2{|\alpha_i|}^2}\left[\langle\hat{a}_s^{\dagger2}\hat{a}_s^2\rangle s_+(g_{sp})+2\bigl\langle\hat{N}_s\bigr\rangle g_{sp}\cos{g_{sp}}\right]-\frac{\operatorname{Var}\hat{N}_s-\bigl\langle\hat{N}_s\bigr\rangle}{4{|\alpha_i|}^2}\sin^3{g_{sp}}\cos{g_{sp}}\left(\sin{4g_{sp}}-\sin{2g_{sp}}-2g_{sp}\right),
\end{split}\end{aligned}
\end{gather}\end{subequations}
where $t_{\mathrm{opt}}$ represents the optimal interaction time that corresponds to the optimal gain $g_{ps} (g_{sp})$ for pump-to-signal (signal-to-pump) transfer.
\end{widetext}

\bibliography{Theory}

\end{document}